\documentclass[aps,prb,twocolumn,showpacs,floatfix,letterpaper]{revtex4}
\usepackage{epsfig}
\usepackage{color}
\usepackage{ulem}
\usepackage{hyperref}

\newcommand{\be}{\begin{equation}}
\newcommand{\ee}{\end{equation}}
\newcommand{\bea}{\begin{eqnarray}}
\newcommand{\eea}{\end{eqnarray}}

\newcommand{\eqref}[1]{(\ref{#1})}

\renewcommand{\k}{{\bf k}}

\begin{document}

\title{Trionic phase of ultracold fermions in an optical lattice: A
  variational study}

\author{\'Akos  Rapp$^{1,2}$, Walter Hofstetter$^3$, and Gergely Zar\'and$^{1,2}$}
\affiliation{
$^1$ Theoretical Physics Department, Institute of Physics, Budapest University of Technology and Economy,
Budapest, H-1521, Hungary,\\
$^2$ Institut f\"ur Theoretische Festk\"orperphysik,Universit\"at Karlsruhe, D-76128 Karlsruhe, Germany,\\
$^3$  Johann Wolfgang Goethe-Universit\"at, D-60438 Frankfurt am Main, Germany
}

\date{\today}

\begin{abstract}

 To investigate ultracold fermionic atoms of three internal states (colors) 
in an optical lattice, subject to strong attractive interaction, 
we study the attractive three-color Hubbard model in infinite dimensions by using a variational
approach. We find a quantum phase transition between a weak-coupling superconducting phase and a strong-coupling trionic phase where groups of three atoms are bound to a composite fermion.
We show how the Gutzwiller variational theory can be reformulated in terms of an effective 
field theory with three-body interactions and how this effective field
theory can be solved exactly in infinite dimensions by using the methods of dynamical mean field theory. 

\end{abstract}


\maketitle

\section{Introduction}
\label{sec:intro}

The Bose-Einstein condensation (BEC) was realized in atomic traps in 1995 
by cooling down $^{87}$Rb atoms in a magnetic trap \cite{Anderson}. 
Since then, experiments have been performed on a variety of ultracold alkali 
atoms ranging from Li to Cs.\cite{Anderson,Li-7,Na-23,Li-6,K-40,Cs-133} 
In addition to displaying new phenomena such as the BEC-BCS crossover\cite{K-40, BEC_BCS}
and the bosonic Mott-transition\cite{bosonic_Mott}, these systems provide 
also clean and flexible realizations for basic theoretical models such as the Hubbard model.\cite{Hubbard_approx_Jaksch, Hubbard_approx_Walter}

Alkali metal isotopes with odd (even) number of neutrons 
behave as fermions (bosons).\cite{Li7+Li6} Although initial experiments were mainly done on bosonic systems, the degenerate Fermi systems have also been realized and confined to optical lattices in recent years due to advanced sympathetic cooling techniques. \cite{evap_cooling,Fermi_deg_opticallattice_Kohl,Fermi_deg_opticallattice_Folling} Because of the Pauli exclusion principle, these fermionic atoms 
can display a variety of interesting phenomena and phases that do not have an analog
in bosonic systems.\cite{Hubbard_approx_Walter}
In the remainder of the paper we shall 
focus on  fermionic systems of three internal degrees of freedom and show how
a quantum phase transition appears in this system, which is a simplified
version of the color superconductor-baryon phase transition in quantum
chromodynamics (QCD). 

Typical hyperfine couplings are larger than the standard experimental
temperatures used to study ultracold gases. In the absence of an external 
magnetic field, the hyperfine coupling aligns nuclear ($\mathbf{I}$) and electronic ($\mathbf S$) spin
antiparallel to each other, and the hyperfine spin $\mathbf F = \mathbf S + \mathbf I$ is
conserved. In finite magnetic fields, only 
the hyperfine spin $F_z$ along the external field 
is a good quantum number. The hyperfine spin $F_z$ thus provides an internal
quantum number that we shall refer to as ``color'' henceforth. 

\begin{figure}[hbp]
\centering
\includegraphics[width=4cm]{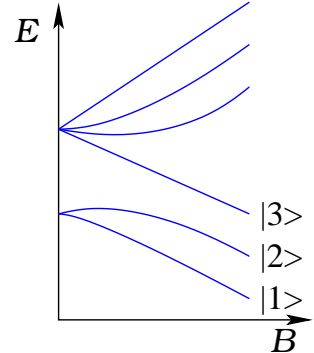}
\caption{\label{fig:Li-hf} (Color online) Sketch of the lowest lying hyperfine levels of $^6$Li in external static magnetic fields. The high-field seeker states have negative tangent, which cannot be trapped in magnetic traps. However, using all-optical setups, one can trap the hyperfine states denoted as $|1 \rangle, |2 \rangle $, and $|3 \rangle$, and thus creating a three-component quantum degenerate fermionic system. }
\end{figure}

Systems with three internal quantum numbers are of special interest,
since they are rarely observed in solid state physics.
Such a three-component fermionic system may be created, e.g., by trapping the lowest
three hyperfine levels of $^6$Li atoms in all-optical setups in large magnetic
fields (see Fig.~\ref{fig:Li-hf}). 

 Such three-component systems with weak interactions have been studied first in 
Ref.~\onlinecite{su3-smallU}, where it has been shown that for small attractive interactions, a color superfluid state emerges. This work has been generalized to incorporate three-body correlations
at large attractive interaction strengths in Ref.~\onlinecite{su3-results-prl}. At the same time, results for
the three-fermion problem in a single parabolic well and a mean field calculation to describe a
two-component BEC-BCS crossover through a Feshbach resonance in $^6$Li appeared.\cite{threefermionproblem,threespeciessuperfluidity}

In this paper, we shall study three-color fermionic systems in an optical
lattice. Optical lattices are realized by standing light waves, which create a periodic potential for the trapped atoms.\cite{Yin} If the amplitude of the lasers is strong enough, then the atoms are localized to
the minima of this potential and at low temperatures can only 
move by tunneling between the lowest lying states within each such minimum.

The trapped atoms, however, also interact with each other. The dominant interactions between alkali atoms are short-ranged electrostatic van der Waals forces, which can be well approximated by $\delta$ potentials at low energy
scales. If the scattering length of the interacting atoms is smaller than the lattice constant of the optical lattice, then these interactions are restricted to a given lattice site. Correspondingly, the interacting system of atoms in an optical lattice can be accurately described by the following Hamiltonian:

\begin{equation}
\hat H = - t \sum_{\langle i,j \rangle, \alpha} {\hat c}_{i\alpha}^+ {\hat
    c}_{j\alpha} + \sum_{\alpha\neq \beta} \sum_i
\frac{U_{\alpha\beta}}{2} 
\left({\hat n}_{i\alpha}{\hat n}_{i\beta}\right) \;, 
\label{eq:H}
\end{equation}
with ${\hat c}_{i\alpha}^+$ the creation operator of a fermionic atom of color $\alpha=1,2,3$ at site $i$, and $\hat n_{i\alpha} =  {\hat c}_{i\alpha}^+ {\hat  c}_{i\alpha}$. In the  tunneling term, $\langle i,j \rangle $ implies restriction to nearest neighbor sites, and the tunneling matrix element is approximately given by $t = E_R (2/\sqrt{\pi}) s^{3/4} e^{-2 s^{1/2}}$, 
where $E_R = \frac{\hbar^2 q^2 }{2m} $ is the recoil energy, $q$ is the wavevector of the lasers,
$m$ is the mass of the atoms, $s= V_0/ E_R$, and $V_0$ is the depth of the 
periodic potential.\cite{Hubbard_approx_Jaksch, Hubbard_approx_Walter} 
We neglect the effects of the confining potential in Eq.~\eqref{eq:H}, which
would correspond to a site-dependent potential term in the Hamiltonian. 
The interaction strength $U_{\alpha\beta}$ between colors $\alpha$ and $\beta$
is related to the corresponding $s$-wave scattering length, $a_{\alpha\beta}$,
as $U_{\alpha\beta} = E_R\; a_{\alpha\beta} q \;\sqrt{8/\pi} s^{3/4}$.\cite{Hubbard_approx_Jaksch,Hubbard_approx_Walter}
Note that fermions with identical colors do not interact with each other.

For the sake of simplicity, we  shall first consider
the  attractive case with $U_{\alpha \beta} = U <0$.  This case could be realized
by loading the $^6$Li atoms into an optical trap in a large magnetic
field, where the scattering  lengths  become 
large and negative $a_{\alpha\beta}\approx a_s\approx -2500\; a_0$, 
for all three scattering channels, $12$, 
$13$, and $23$.\cite{feshbach-Li} 

Introducing the usual Gell-Mann matrices,   $\lambda_{\alpha\beta}^a$  ($a=1,\dots,8$),
it is easy to see that global SU(3) transformations ${\rm exp} ( i \sum_i \sum_{a\alpha\beta} \phi_a \hat c_{i \alpha}^+ \lambda_{\alpha\beta}^a \hat c_{i \beta} )$ 
commute with the Hamiltonian, which thus also conserves the total number of fermions 
for each color,  $\hat N_\alpha = \sum_i  \hat n_{i\alpha}$. 
This conservation of particles is only  approximate because 
in reality  the number of the atoms in the trap continuously 
decreases due to different scattering processes. 
Here, however, we shall neglect this slow loss of atoms and keep the 
density $\rho_\alpha$ of atoms for color $\alpha$ as well as the overall 
filling factor $\rho \equiv \frac 13 \sum_\alpha \rho_{\alpha}$ fixed.

\begin{figure}[htp]
\centering
\includegraphics[width=6cm]{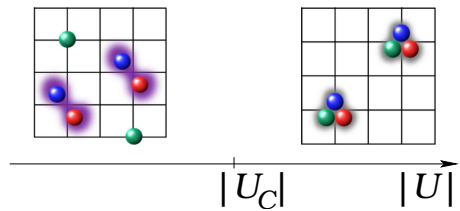}
\caption{\label{fig:pert}(Color online) The ground states for $|U| \ll |U_C|$ and $|U| \gg |U_C|$ can be calculated by perturbation theory. The former is a BCS-state which breaks the SU(3) symmetry, and the latter is a trionic state with three-particle singlet bound states.} 
\end{figure}

\begin{figure}[hbp]
\centering
\includegraphics[width=6cm]{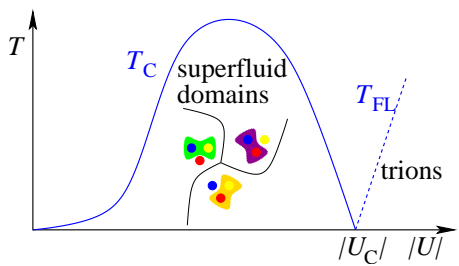}

\caption{\label{fig:sketch}(Color online) Schematic phase diagram of the attractive SU(3) Hubbard model. The color superfluid phase is stable below a critical temperature $T_C$, where the appearance of superfluid domains is expected. At large interaction strengths $|U| > |U_C|$, color singlet trions emerge instead of the Cooper pairs and form a Fermi liquid at low temperatures $T < T_{FL}$.  }
\end{figure}

Let us first focus on the case of equal densities, $\rho_\alpha=\rho$.
For small attractive $U<0$, 
the ground state is a color superfluid:\cite{su3-smallU}
atoms from two of the colors form the Cooper pairs and an $s$-wave 
superfluid, while the third color 
remains an unpaired Fermi liquid. However, as we discussed in Ref.~\onlinecite{su3-results-prl}, for large
attractive interactions, this superfluid state becomes unstable, and
instead of Cooper pairs, it is more likely to form three-atom bound states, the
so-called ''trions". These trions are color singlet
fermions, and for large $|U|$ they have a hopping amplitude,
\begin{equation}
t^{\rm trion} \sim  \frac{t ^3}{4U^2}. 
\end{equation}
Furthermore, one can easily see that if two trions sit on neighboring
lattice sites, then they increase the energy of each other
by an amount $V\sim t^2/(2\vert U\vert)$. This is because 
the energy of an individual trion is decreased by quantum fluctuations
where one of the atoms virtually hops to one of the neighboring sites. 
These quantum fluctuations are reduced if the two trions sit next to 
each other. Therefore, trions will tend to form a Fermi-liquid in any finite dimensions.
This Fermi liquid state may be further
decorated by charge density wave order at large values of $|U|$. Also, 
the Fermi liquid scale $T_{FL}$ of the trionic 
Fermi liquid should depend on the value of $U$, and at the 
transition point, $U=U_C$, we expect it to go zero (see Figs.~\ref{fig:pert}
and \ref{fig:sketch}).

In order to get analytic expressions, we shall 
study the ground state in $d= \infty$ dimensions. 
Then, to reach a meaningful limit and to get finite kinetic energy, 
one has to scale the hopping as $t= \frac{t^*}{2\sqrt{d}}$, 
with  $t^*$  fixed. In this limit, however, trions become 
immobile. Therefore the $d\to\infty$ trionic states are well approximated 
as 
\begin{equation} 
\vert T_{\Lambda} \rangle = \prod_{i \epsilon \Lambda}  \hat c_{i1}^+
\hat c_{i2}^+ \hat c_{i3}^+ \vert 0\rangle \;, \label{eq:def_trionic_state}
\end{equation} 
where $\Lambda$ denotes a subset of sites where trions sit. We can calculate the
energy of this state in infinite dimensions: a single trion has an energy
$3U$, thus the energy of such a state per lattice site is given 
by $E_T/N = 3 U \rho$, with $E_T$ the total energy of the system and 
$N$ the number of lattice sites.

Clearly, the two ground states obtained by the perturbative expansions have different
symmetries: the superfluid state breaks SU(3) invariance, while the
trionic state does not. Therefore, there must be a phase transition between them. 
Note that, relying on symmetries only, this argument is very robust and
carries over to \emph{any} dimensions. In infinite dimensions, we find that trions are immobile. However, this is only an artifact of infinite dimensions and in finite dimensions, a superfluid-Fermi liquid phase transition should occur.

One could envision that some other order parameter also emerges
and masks the phase transition discussed here. 
Preliminary results (not discussed here) suggest that indeed a charge density state forms at large
values of $|U|$, but except for half filling, which is a
special case not discussed here, we do not see any other relevant order
parameter that could intrude as a new phase. One could, in principle, also
imagine a phase with  simultaneous trionic and fermionic Fermi
surfaces, similar to the one of Ref.~\onlinecite{Subir}. However, in
contrast to the calculations of Ref.~\onlinecite{Subir} (which do not take the  
local constraints on the lattice into account), in our scenario, there is no Fermi
surface at the quantum critical point.

From the above argument, it is unclear whether the phase transition 
is of first or of second order. To address this question in infinite 
dimensions and give some more quantitative estimates for 
the relevant condensation energies and the critical value of 
$U/t$, we need an approach that is able 
to describe the  superfluid state and, at the same time, also accounts for three-body correlations. 
In the present paper, we construct a variational Gutzwiller wave function
that is able to capture both correlations simultaneously. We then 
show how averages can be evaluated by constructing an effective 
action that contains three-body correlations and 
how calculations can be analytically done in infinite dimensions by
using the methodology of dynamical mean field theory.\cite{dmft-revmodphys}
Our method, which uses a single suitably chosen Gutzwiller
correlator and exploits the cavity method of dynamical mean field theory, 
is in its present form less flexible than the multiband method reviewed by B\"unemann \textit{et. al.,}\cite{multiband_gutzwiller} but it incorporates three-body
correlations in a very transparent way. Our analysis shows that, within the Gutzwiller approach, the phase transition is of {\em second order} in infinite dimensions. We therefore 
expect it to remain of second order in any finite dimension above 
the lower critical dimension. Nevertheless, as we discuss later, the
quantum phase transition may become of slightly first order 
due to a not perfectly SU(3)-symmetrical interaction.

Interestingly, our results also suggest that there is 
a tendency to create an imbalance of the densities 
$\rho_\alpha$ in the superfluid phase. The physical reason is simple: 
one can gain condensation energy by transferring atoms  
from the unpaired channel to the paired ones and thereby create 
{\em ferromagnetic order} as a secondary order parameter.
For equal densities $\rho_\alpha = \rho$ and color conservation, his can only happen if the atoms are segregated and domains are formed, as shown in Fig.~\ref{fig:sketch}, where we sketched 
the schematic phase diagram of the attractive SU(3)  
model away from half-filling. This picture, which is first proposed in
Ref.~\onlinecite{su3-results-prl}, has also been confirmed by the Ginzburg--Landau-type effective field theoretical approach of Ref.~\onlinecite{segregation_Cherng}.

So far, we discussed the fully SU(3) symmetrical case. 
As we also demonstrate later, the phase transition discussed 
above is not sensitive to having perfect SU(3) symmetry. On the other hand, 
to form the superfluid phase, it is important to have 
approximately the same Fermi momenta for at least 
two colors.

The rest of the paper is organized as follows. In Sec. \ref{sec:Gutzw_ans}, we
introduce the Gutzwiller ansatz for the ground state. In Sec. \ref{sec:Eff_theo}, we
reformulate the Gutzwiller expectation values as an effective path
integral. In Sec. \ref{sec:cavity}, we derive a local action, which can be used to solve
the effective action in the $d=\infty$ dimensional limit. In Sec. \ref{sec:solution}, we
first summarize the results for the SU(3) symmetric case and then generalize the approach by analyzing the
Hamiltonian with nonuniform interaction strengths in order to describe a system
of $^6$Li atoms. In Sec. \ref{sec:qcd}, we present a brief discussion of analogies
with QCD. In Sec. \ref{sec:conclusions}, we present our conclusions. Some
of the technical details can be found in the Appendices.

\section{Gutzwiller ansatz}
\label{sec:Gutzw_ans}

To capture the color superfluid - trion transition, we approximate the ground state of the infinite-dimensional system by the following Gutzwiller-correlated  wave function:
\begin{equation}
\vert G \rangle = \prod_i [1+ (g-1) \hat n_{i1} \hat n_{i2} \hat n_{i3} ]
\vert {\rm BCS} \rangle \;. 
\label{eq:def_gutzwillerstate} 
\end{equation}
Here $g$ is a Gutzwiller variational parameter, which increases (or decreases) the amplitude of terms in the ``uncorrelated-state'' which have triple occupancies. In the $g\to \infty$ limit, $\vert G \rangle$ becomes  a
superposition of trionic states similar to Eq.~(\ref{eq:def_trionic_state}). We
choose the uncorrelated ground state as a BCS-like state with  colors  ``1"
and ``2'' forming Cooper pairs,\cite{SU3-rotation}
\begin{equation}
\vert {\rm BCS} \rangle = \prod_{\k:\; \epsilon_\k < \mu_3} \hat c_{\k 3}^+
\prod_{\k '} (u_{\k'} + v_{\k'}  \hat c_{\k' 1}^+ \hat c_{-\k' 2}^+ ) \vert 0
\rangle \;. 
\label{eq:def_bcsstate} 
\end{equation}
The operators $\hat c_{\k \alpha}^+$  in Eq.~(\ref{eq:def_bcsstate}) diagonalize the first part of the Hamiltonian (\ref{eq:H}), and 
create  fermions with momentum  $\k$, color $\alpha$, and 
energy $\epsilon_\k=-t\sum_{\bf a}
e^{i\k \mathbf a}$, with the vector $\bf a$ running over nearest neighbor
sites. The coefficients 
 $u_\k^2=\frac{1}{2} ( 1 + \xi_\k / \sqrt{\xi_\k^2 + \Delta^2} )$ and
$v_\k =
\sqrt{1 - u_\k^2}$ are the usual BCS coherence factors, with $\xi_\k =
\epsilon_\k - \mu_{12}$.  The ``chemical potentials''
$\mu_3$ and  $\mu_{12}$ appearing in the wave function 
should be considered merely as parameters that are adjusted to fix the densities
$\rho_\alpha$ at a given value. 

To perform a variational calculation, we need to evaluate the Gutzwiller expectation value
of Hamiltonian (\ref{eq:H}),
\begin{equation} 
\langle \hat H \rangle_{G} = \langle G \vert \hat H \vert G \rangle / \langle G \vert G \rangle \;,
\end{equation} 
at a given filling factor $\rho$, and then minimize it with respect to $g$,
$\Delta$, and eventually the density of the third color, $\rho_3$.

We note that our variational wave function  smoothly interpolates between the color superfluid $(g=1)$ and the trionic state $(g\to\infty)$. We can also compare the Gutzwiller energy to certain reference states, such as the Fermi sea ($g=1$, $\Delta=0$), the Gutzwiller correlated Fermi sea ($g\ne1$, $\Delta=0$), or the uncorrelated BCS state ($g=1$, $\Delta\ne0$), and thereby estimate correlation or condensation energies.

\section{Effective Grassmann theory}
\label{sec:Eff_theo}

We first derive an effective action that can be used to replace the Gutzwiller expectation value of an operator $\hat O$,
\begin{equation}
\langle \hat O \rangle_G = \langle G \vert \hat O \vert G \rangle / \langle
G \vert G \rangle, 
\label{eq:def_Gutzwiller_expval} 
\end{equation}
by a combination of the Grassmann path integrals. For this purpose, let us first rewrite the denominator of Eq. (\ref{eq:def_Gutzwiller_expval}) as 
\begin{equation}
\langle G \vert G \rangle = \langle  {\rm BCS}  \vert \prod_i [1+(g^2 -1 ) \hat n_{i1}
\hat n_{i2} \hat n_{i3} ] \vert  {\rm BCS}  \rangle. \label{eq:normalorder_GG} 
\end{equation}
Here, the operator in the middle is in a normal ordered form (i.e., all $\hat
c^+_{i\alpha}$ appear to the left of the operators $\hat c_{i\alpha}$). Since
the BCS state is a non-interacting state (in terms of bogoliubons), 
we can use Wick's theorem to evaluate expectation values
appearing in Eq.~(\ref{eq:normalorder_GG}). Due to the normal ordering, every $\hat c^+_{i\alpha}$
and $\hat c_{i\alpha}$ appears in Eq.~(\ref{eq:normalorder_GG}) only once. Therefore, we can calculate the quantum mechanical expectation values as the Grassmann path integrals, where we replace 
normal ordered operators by the Grassman variables as 
 $\hat c_{i\alpha}\to \eta_{i\alpha}$ and $\hat c_{i\alpha}^+\to \bar
 \eta_{i\alpha}$ and evaluate expectation values with an appropriately chosen  
action ${\cal S}_0$, 
\begin{eqnarray}
&\langle :\cdots: \rangle_{BCS} \to 
\langle \cdots \rangle_ {{\cal S}_0 } \equiv 
\int {\cal D} \bar \eta {\cal D} \eta \cdots e^{-{\cal S}_0}, 
\nonumber \\
&{\cal S}_0 = - \frac{1}{2} \sum_{ij} \bar \Psi_i [D^{0}]^{-1}_{ij} \Psi_j . 
\end{eqnarray}
Here, we introduced the ``Nambu spinors'', $\bar \Psi_i = (\bar \eta_{i\alpha},
\eta_{i\alpha})$, 
and the propagator $D^{0}_{ij}$ must be chosen so that it satisfies the conditions,
\begin{eqnarray}
\langle \bar \eta_{j\beta} \eta_{i\alpha}\rangle_ {{\cal S}_0 }&\equiv&
\langle \hat c_{j\beta} ^+ \hat c_{i
  \alpha}  \rangle_{BCS} \equiv G^0_{ij \alpha \beta }, 
\nonumber
\\
\langle \eta_{j\beta} \eta_{i\alpha}\rangle_ {{\cal S}_0 }
&\equiv&  
\langle  \hat c_{j\beta} \hat c_{i \alpha}  \rangle_{BCS}  \equiv F^0_{ij \alpha \beta} ,   
\nonumber
\end{eqnarray}
with $G^0$ and $F^0$ the  normal and anomalous Green's functions. 
It is easier to express $D^{0}$ in the Fourier space, where it is just
a $6\times6$ dimensional matrix in the Nambu space that can be expressed as
\begin{equation}
D^0 (\k) = \left(\begin{array}{cc} G^0(\k) & F^0(\k) \\ F^{0 +}(\k) &
      -G^{0 +}(\k) \end{array} \right).
\end{equation}
We can thus rewrite $\langle G \vert G \rangle $ as 
\begin{equation}
\langle G \vert G \rangle = \left\langle \prod_i \left[  1+(g^2 -1 )\prod_\alpha
\bar \eta_{i\alpha} \eta_{i\alpha} \right] \right\rangle_{{\cal S}_0} \;. 
\label{eq:grassman_0}
\end{equation}
Note that in the above procedure, we have not doubled the Hilbert space since the
integration is over the fields, $\bar \eta$ and $\eta$, and therefore the $\Psi$ and $ \bar \Psi$ do not represent independent Grassman variables. 

Note also that the Green's functions
$ G^0_{ij \alpha \beta }  = \langle  \hat c_{j\beta} ^+ \hat c_{i \alpha} \rangle_{BCS}$ 
and 
$F^0_{ij \alpha \beta} =  \langle  \hat c_{j\beta} \hat c_{i\alpha} \rangle_{BCS}$ 
can easily be determined in terms of the BCS coherence factors. 
Furthermore, neither the Green's functions 
nor the Grassman fields $\eta_{j\beta}$  have a time 
argument. These Grassman fields should thus not be confused with the 
Grassman fields that appear in the path integral expression of the density
matrix or the time evolution operator.

To make further progress, let us define the ``triple occupancy'' as $t_i \equiv n_{i1}n_{i2}n_{i3}  \equiv \prod_\alpha \bar \eta_{i\alpha} \eta_{i\alpha}$. Using the properties of the Grassmann variables, this quantity
can be expressed as a function of the Grassmann vectors defined above
\begin{equation}
 t_i = \frac{1}{48} \left(\bar \Psi_i  \tau_3 \Psi_i \right)^3 ,
\end{equation}
where $\tau_3 = \sigma_3 \otimes \delta_{\alpha \beta}$, and the third Pauli matrix
 $\sigma_3$ acts in Nambu space. Clearly, $t_i^2 = 0$. Therefore, exponentiating the product in Eq.~(\ref{eq:grassman_0}), we obtain the relation
\begin{equation}
\langle G \vert G \rangle = \int {\cal D} \bar \eta {\cal D} \eta \; e^{-{\cal S}}, 
\end{equation}
where the full (interacting) effective action is given by
\begin{equation}
{\cal S} = - \frac{1}{2} \sum_{ij} \bar \Psi_i [D^{0 -1}]_{ij} \Psi_j -
(g^2-1) \sum_i t_i \;. \label{eq:def_action}
\end{equation}
Interestingly, the Gutzwiller variational parameter appears in this effective
action as a three-body interaction. For the uncorrelated wave function, 
$g=1$, the interaction term vanishes. For trionic correlations one has 
$g>1$, i.e., the effective three-body interaction is attractive, while for 
$g<1$ the effective interaction is repulsive.

This procedure can be repeated with certain operator's quantum mechanical
expectation values after proper normal ordering. For local operators $\hat
O_i$, the Gutzwiller expectation value is 
\begin{eqnarray}
&&{\langle G \vert \hat O_i \vert G \rangle}/{\langle G \vert G \rangle} =
\nonumber \\ 
&=&{\langle {\rm BCS} \vert :\hat O_i:_G \prod_{m\neq i} [1+(g^2-1)
  \hat t_{m} ]  \vert {\rm BCS} \rangle}/{\langle G \vert G \rangle} \;,\nonumber 
\end{eqnarray}
where the Gutzwiller normal ordered operator $:\hat O_i:_G$ is defined as 
follows:
\begin{equation}
 :\hat O_i:_G = :[1+(g-1)\hat t_{i} ]\hat O_i [1+(g-1)\hat t_{i}]:
\end{equation} 
and $\hat t_i = \hat n_{i 1} \hat n_{i 2} \hat n_{i 3}$. Here, the numerator can also be converted into expectation values evaluated with $\cal S$: We first use Wick's
theorem and replace all operators by the Grassman variables and the average 
by $\langle\dots\rangle_{{\cal S}_0}$. Then, we can insert 
the missing term $[1+(g^2-1) t_{i} ]$ and reexponentiate 
the product to have an average with the action $\cal S$. 

With this procedure, the particle density and double occupation 
on a site can be written as follows:
\begin{eqnarray}
N_i &=& \langle \hat n_{i1} + \hat n_{i2} + \hat n_{i3} \rangle_G \nonumber \\
&=& \langle n_{i1} + n_{i2} + n_{i3} \rangle_{\cal S} + 3 (g^2 - 1)
\langle t_i  \rangle_{\cal S},
\\
D_i &=& \langle \hat n_{i1} \hat n_{i2} + \hat n_{i2} \hat n_{i3} +  \hat
n_{i1} \hat n_{i3} \rangle_G \nonumber \\ &=&  \langle n_{i1}n_{i2}  +
n_{i2}n_{i3}  + n_{i1} n_{i3} \rangle_{\cal S} \nonumber \\
&&+ 3 (g^2 - 1)  \langle t_i
\rangle_{\cal S}. 
\end{eqnarray}
To evaluate the expectation value of the kinetic energy, we also need
to evaluate expectation values of the type $\langle \hat c_{i\alpha}^+ \hat c_{j\alpha} \rangle_G$ for $i\ne j$. 
We therefore define the density supermatrix (formulas here are valid for $i\neq j$), 
\begin{equation}
{\cal P}_{ij} = \left(\begin{array}{cc} P_{ij} & Q_{ij}^+ \\ Q_{ij} & - P_{ij}^+ \end{array} \right),
\end{equation}
with the $3\times3$ matrices $P_{ij}$ and $Q_{ij}$ defined as
\begin{eqnarray}
P_{i\ne j,\alpha\beta} &=& \langle \hat c_{i\alpha}^+ \hat c_{j \beta} \rangle_G = \langle \bar \eta_{i\alpha} \eta_{j\beta} \rangle_{\cal S} \nonumber \\
&& -(1-g)\langle \bar \eta_{i\alpha} \eta_{j\beta}(d_{i\alpha} + d_{j\beta})\rangle_{\cal S} \nonumber \\
&& + (1-g)^2\langle \bar \eta_{i\alpha} \eta_{j\beta}(d_{i\alpha} d_{j\beta} ) ) \rangle_{\cal S} \; 
\label{eq:P_ij}
\end{eqnarray}
and
\begin{eqnarray}
Q_{i\ne j,\alpha\beta} &=& \langle \hat c_{i\alpha} \hat c_{j \beta} \rangle_G = \langle \eta_{i\alpha} \eta_{j\beta} \rangle_{\cal S} \nonumber \\
&& -(1-g)\langle \eta_{i\alpha}
\eta_{j\beta}(d_{i\alpha}  + d_{j\beta})\rangle_{\cal S} \nonumber \\ 
&& + (1-g)^2 \langle \eta_{i\alpha} \eta_{j\beta} d_{i\alpha} d_{j\beta}
\rangle_{\cal S}.\label{eq:Q_ij} 
\end{eqnarray}
In these expressions, we defined the Grassman factors 
$d_{i\alpha} \equiv  \prod_{\beta \ne \alpha }n_{i\beta}$. 
Clearly, for $g=1$, the matrix ${\cal P}$ is simply related to the 
unperturbed Green's function, ${\cal P}(g=1) = [D^0]^T $. As we shall now
see, although the above expressions look rather complicated and contain
formally three-body correlation functions in terms of the effective field
theory,  we shall still be able to express 
 $\cal P$ in terms of the full single particle Green's function 
$D$, defined as 
\begin{equation} 
D_{ij} \equiv - \langle \Psi_i \bar \Psi_j \rangle_{\cal S} =
\left(\begin{array}{cc} G_{ij} & F_{ij} \\ F^{+}_{ij} & -G^{+}_{ij}
  \end{array} \right),
\end{equation}
and its proper self-energy $\Sigma$. 

\begin{figure}[bp]
\centering
\includegraphics[width=6cm]{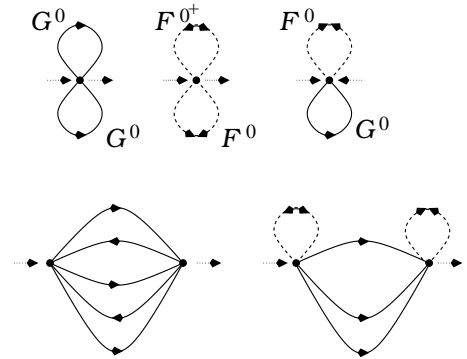}
\caption{\label{fig:self-energy} Examples of first and second order self-energy
diagrams. Continuous lines denote the unperturbed normal Green's functions, the staggered lines the anomalous ones, while vertices stand for the three-body interaction, $\sim (g^2-1)$. In infinite dimensions, we shall sum up these diagrams by using a \emph{nondiagrammatic} method.}
\end{figure}

At this point, it is useful to define the generating functional,
\begin{equation}
\Gamma = \ln \int {\cal D} \bar \eta {\cal D} \eta {\rm exp} \left( -{\cal S} + \sum_i \bar I_i
  \tau_3 \Psi_i \right),\label{eq:def_gamma}
\end{equation}
where the ``current''  $\bar I_i$ is a six-component Grassmann vector,
$\bar I_i = (\bar J_{i1},\bar J_{i2},\bar J_{i3},J_{i1},J_{i2},J_{i3})$, 
and make use of standard field theoretical methods.\cite{Negele} 
As usual, the components of the Green's function can be obtained from
$\Gamma$ by functional derivation,
\begin{equation}
\begin{array}{cc} G_{ij\alpha\beta} = \frac{ \delta^2 \Gamma}{\delta \bar
    J_{i\alpha} \delta J_{j\beta} } \Bigg \vert_{\bar I =0}     &
  F_{ij\alpha\beta} = - \frac{ \delta^2 \Gamma}{\delta \bar J_{i\alpha} \delta
    \bar J_{j\beta} } \Bigg \vert_{\bar I =0}  \end{array} .
\end{equation}
The (improper) self energy $S$ is defined by cutting off the bare lines from the dressed propagator:
\begin{equation}
D_{ij} = D_{ij}^0 + \sum_{pq} D_{ip}^0  S_{pq} D_{qj}^0,
\;,\label{eq:def_selfenergy}
\end{equation}
while the proper (one particle irreducible) self-energy obeys the 
Dyson's equation,
\begin{equation}
S_{ij} = \Sigma_{ij} + \sum_{pq}  \Sigma_{ip} D_{pq}^0 S_{qj}  \;.
\label{eq:Sigma_S}
\end{equation}
It is useful to visualize these self-energies in terms of the
Feynman diagrams of the effective field theory, as shown in Fig.~\ref{fig:self-energy}, although we shall sum up these diagrams later by using nondiagrammatic methods. As mentioned earlier, the Gutzwiller correlator 
generated a three-body interaction in the effective field theory, while
anomalous Green's functions appear due to the superfluid correlations.

To generate the necessary identities to evaluate the expectation values in 
Eqs.~(\ref{eq:P_ij}) and (\ref{eq:Q_ij}), we shift the Grassmann fields $\bar \eta_i$ and $\eta_i$ in the generating functional $\Gamma$ as
\begin{equation}
\Psi_i \to \Phi_i = \Psi_i + \lambda \sum_p   D^{0}_{ i p } \tau_3 I_p,\label{eq:def_trf}
\end{equation}
where $\lambda$ is a real valued parameter. Note that 
this equation also determines how  the  field $\bar \Psi_i $ is changed,
since the components of $\bar \Psi_i $ and $ \Psi_i $ are related.
Proceeding with the functional derivation after this change of integration 
variables leads to an expression of the dressed propagator $D$ in terms  
of expectation values of the Grassmann fields related to the expectation values 
in Eqs.~(\ref{eq:P_ij}) and (\ref{eq:Q_ij})  (for details, see Appendix \ref{app:identities}.),
and finally comparison of these with the definition of the self-energy gives the following identities ($i \neq j$): 
\begin{eqnarray}
&-& (g^2 - 1) ^2 \tau_3  \langle \Psi_i d_i^F d_j^F \bar \Psi_j \rangle_{\cal S} \tau_3 = S_{ij},\label{eq:identities-1}\\ 
&&(g^2 - 1) \langle \Psi_i d_j^F \bar \Psi_j \rangle_{\cal S} \tau_3 =  \sum_p D^0_{ip} S_{pj},\label{eq:identities-2}\\
&&(g^2 - 1) \tau_3 \langle \Psi_i d_i^F \bar \Psi_j \rangle_{\cal S} =  \sum_p S_{ip} D^0_{pj},
\label{eq:identities-3}
\end{eqnarray}
where $d_i^F = d_{i1} + d_{i2} + d_{i3} =  \frac{1}{8} (\bar \Psi_i  \tau_3 \Psi_i )^2 $.
Using these identities, it is possible to express the density supermatrix 
$\cal P$
in terms of the improper self-energy. This relation is more transparent in the Fourier space, where it reads
\begin{eqnarray}
{\cal P}^t (\k) &=& D^0(\k)  + C \nonumber \\
 && + \left( D^0(\k) - \frac{
 \tau_3}{g+1} \right) S(\k)  \left( D^0(\k) - \frac{ \tau_3}{g+1} \right), \nonumber\\ \label{eq:calP_S}
\end{eqnarray}
where the superscript $t$ refers now to the transposed matrix in color space, and $C$ is a $\k$-independent $6\times6$ matrix related to the $i=j$ contributions. This term does not play a role in the evaluation of the kinetic energy,
\begin{equation}
K = \sum_\k  \epsilon_\k \sum_\alpha P_{\alpha\alpha}(\k),
\end{equation}
since we assumed only nearest neighbor hopping, and therefore $\sum_{\k,\alpha}
C_{\alpha\alpha} \epsilon_\k = 0$.

\section{Cavity functional and connection to Dynamical Mean field Theory}
\label{sec:cavity}

In the previous section, we showed that the Gutzwiller expectation values can be
expressed by path integrals, which still need to be computed. These path
integrals cannot be exactly evaluated in general. However, an important
simplification occurs in the limit of infinite dimensions, $d\to \infty$. There the off-diagonal ($i\neq j$) Green's functions decay as $D_{ij}\sim 1/d^{||i-j||/2}$, where  $||i-j||$ denotes the minimum number of
steps that are needed to reach the lattice point $i$ from the other lattice
point $j$.\cite{dmft-revmodphys} Therefore, the
off-diagonal ($i\neq j$) components of the proper self energy 
rapidly go to zero, $\Sigma_{ij}\sim 1/d^{3 ||i-j||/2}$, and vanish in the 
infinite-dimensional limit, where the self-energy becomes local,
\begin{equation}
\Sigma_{ij} = \delta_{ij} \Sigma \;.
\end{equation}
Here, we already used the discrete translational symmetry of the lattice:
$\Sigma(i) = \Sigma$. 

As $\Sigma$ is a central quantity of interacting systems and allows one to compute virtually any other quantity, our main goal shall be to determine $\Sigma$. This step is done similar to the cavity method also used in dynamical mean field theory to extract local properties at a mean field level.\cite{dmft-revmodphys} In this approach one integrates out the Grassmann variables on all sites except for the origin,
\begin{equation}
\frac{1}{Z} \int {\cal D}' \bar \eta {\cal D}' \eta  e^{-{\cal S}} = \frac{1}{Z_L}
e^{-{\cal S}_L} \;,
\end{equation}
where the prime indicates that one should not integrate over the variables 
$\eta_0$ and $\bar \eta_0$  at the origin. 
Thus, by construction, the local action depends only on the variables $(\bar \eta_0, \eta_0)$, 
${\cal S}_L = {\cal S}_L[\bar \eta_0, \eta_0]$, and all local expectation values
and correlation functions are invariant,
\begin{equation} 
\langle O_0 \rangle_{\cal S} = \langle O_0 \rangle_{{\cal S}_L}
\;. \label{eq:selfcons000}
\end{equation}
Importantly, the local Green's function is also invariant. 

Since the hopping is scaled as $\sim 1/\sqrt{d}$, it can be shown (Appendix
\ref{app:localaction}) that in infinite dimensions, the generated local action
takes on a simple functional form,
\begin{equation}
-{\cal S}_L = \frac{1}{2} \bar \psi {\cal D}_0^{-1} \psi  + 
\frac{g^2-1}{48} (\bar  \psi \tau_3 \psi)^3 \;,
\label{eq:S_L}
\end{equation}
where ${\cal D}_0$  denotes the ``cavity Green's function''. The field $\psi$ in Eq.~(\ref{eq:S_L})
denotes the Grassmann field $\Psi_i$ in the origin,  $\psi \equiv\Psi_0$. 

The cavity Green's function ${\cal D}_0$ can be determined selfconsistently 
from the condition that Eq. (\ref{eq:selfcons000}) holds for any local quantity, 
including the local dressed propagator of the effective lattice theory,
\begin{equation}
D^{\rm loc} \equiv D_{00} = \sum_\k D(\k),
\end{equation}
which, by construction, must coincide with the full Green's function of the
cavity theory, 
\begin{equation} 
D^{\rm loc}= {\cal D} = - \langle \psi  \bar \psi \rangle_{{\cal S}_L}.
\end{equation}

Then, on one hand, the local propagator on the lattice can be determined by using
Dyson's equation and thus can be expressed in terms of 
$\Sigma$, as 
\begin{equation} 
D^{\rm loc} = \sum_\k [D^{0} (\k)^{-1} - \Sigma]^{-1}\;.
\label{eq:dyson_local_latt}
\end{equation}
On the other hand, we can also compute the local propagator {\em exactly} by 
evaluating the path integral with Eq.~(\ref{eq:S_L}) analytically: 
Using symmetry considerations presented in Appendix \ref{app:diagonal}, 
we can show that there is  a $\cal U$ unitary transformation that diagonalizes ${\cal D}_0$:
\begin{equation}
{\cal U} {\cal D}_0 {\cal U}^+ =\left (\begin{array}{cc} {\rm d} & 0 \\ 0 &
    -{\rm d} \end{array} \right),
\label{eq:U}
\end{equation}
where  $\rm d$ is a $3\times3$ (real) diagonal matrix. We  can then evaluate the
dressed propagator in the local theory by performing  a variable
transformation $\psi \to \phi = {\cal U} \psi$ in the intergal, the result being 
\begin{equation}
D^{\rm loc} ={\cal D} = \frac{ {\cal D}_0}{1 + (g^2 - 1) \sqrt{- \det {\cal D}_0}} .
\label{eq:sc}
\end{equation}
Furthermore, Dyson's equation also holds for the local theory, and moreover 
in infinite dimensions, the proper self-energy $\Sigma$ of the lattice theory
is the same as in the self-energy in the local theory. This follows from 
the comparison of the skeleton diagrams of the local and lattice theories, and
the fact that the full Green's functions are the same in both theories.
We can thus write 
\begin{equation}
\Sigma = {\cal D}_0^{-1}-{\cal D}^{-1} = -(g^2-1) \sqrt{- \det {\cal D}_0} {\cal D}_0^{-1}.
\label{eq:dyson_local}
\end{equation}

With some algebraic manipulations, Eqs.~(\ref{eq:dyson_local_latt}) and (\ref{eq:dyson_local}) 
can be recast in the following self-consistency condition,
\begin{equation}
 - \Sigma^{-1} = \left( 1 + \sqrt{\frac{g^2 - 1}{\sqrt{-\det \Sigma}}}
 \right)   \sum_\k [D^{0 -1} (\k) - \Sigma]^{-1} \;.
\label{eq:selfcons_equations}
\end{equation}
This is the central equation of the theory. Having solved this equation self-consistently for the
self-energy matrix $\Sigma$, we can express, e.g., the cavity Green's function as
\begin{equation}
{\cal D}_0 = -  \sqrt{ \frac{ \sqrt{ -\det \Sigma}}{ g^2 - 1} }   \Sigma^{-1} , 
\end{equation}
and then determine from that the local Green's function using 
Eq.~(\ref{eq:sc}) or the improper self-energy from the Fourier
transform of Eq.~(\ref{eq:Sigma_S}). Remember though that the quantities here are $6\times6$ matrices and
inversions mean matrix inversions.

Setting all these together and using the relation [Eq.~(\ref{eq:calP_S})] between the density matrix ${\cal P}_{ij}$
and the improper self-energy, the kinetic energy can be finally expressed as
\begin{eqnarray}
K &=& \sum_\k \epsilon_\k \sum_{\alpha=1}^3 \big[ \frac{}{} [D^{0-1}(\k) -
  \Sigma]^{-1} \nonumber \\
&& - \frac{2}{1+g} [D^{0-1}(\k) - \Sigma]^{-1} \Sigma  \nonumber \\
&& + \frac{1}{(1+g)^2} D^{0-1}(\k) [D^{0-1}(\k) - \Sigma]^{-1} \Sigma \big]  _{\alpha\alpha},
 \nonumber 
\end{eqnarray}
where the summation runs only over the first three diagonal elements of the 
$6\times6$ matrices. We can also compute other local expectation values using the same variable
transformation [Eq.~(\ref{eq:U})] as before. The total particle density is is given by
\begin{eqnarray}
N_T &=& 3 \rho = \frac{1}{2} \langle \bar \psi \tau_3 \psi  
\rangle_{{\cal S}_L} + 3 (g^2-1) \langle t \rangle_{{\cal S}_L} \nonumber \\
&=& \sum_{\alpha=1}^3  \frac{-\Sigma_{\alpha\alpha}^{-1} + 1}
{1 + \sqrt{\frac{g^2 - 1}{\sqrt{-\det \Sigma}}} },  \nonumber 
\\ 
\end{eqnarray}
while the total double occupancy is
\begin{eqnarray}
D_T&\equiv& \sum_{\alpha=1}^3  \langle d_\alpha\rangle_G 
= \frac{1}{8} \langle (\bar \psi \tau_3 \psi)^2 \rangle_{{\cal S}_L} + 3
(g^2-1) \langle t \rangle_{{\cal S}_L}  
\nonumber 
\\ 
&=& \sum_{\alpha=1}^3 \frac{\Sigma_{\alpha\alpha} - \sqrt{(g^2 - 1)\sqrt{-\det \Sigma}} }{1- g^2 - \sqrt{(g^2 - 1)\sqrt{-\det \Sigma}}  }  \;.
\end{eqnarray}

This is a good point to discuss the similarities and differences
with the multiband Gutzwiller method  of B\"unemann \textit{et. al.}\cite{multiband_gutzwiller} That approach employs a flexible Gutzwiller correlator with multiple variational parameters. Some of these
parameters are allowed to be fixed so that the proper self-energy of the
effective theory vanishes in infinite dimensions. This leads to 
important simplifications when calculating the Gutzwiller expectation values.
In our formalism, on the other hand, we used a single correlation parameter that 
was motivated by weak and strong-coupling theories. Nevertheless, this single correlator is able to capture the physics of both limits and it therefore accounts for the most important correlations. Including additional (e.g., two-body correlators) should quantitatively modify our results, and change the critical value of $U$, however, we do not expect qualitative changes due to them. The advantage of our effective field theory approach is that it displays the trionic correlations in a rather transparent way, in terms of three-body interactions. Preliminary results indicate that our cavity functional approach could also be modified to use more flexible correlators, but this is out of the scope of the present paper.

\section{Solutions of the self-consistency equations}
\label{sec:solution}

\subsection{Derivation of integral equations}

The self-consistency equations are hard to solve in general. However, we can 
simplify their solution by observing that the proper self-energy must have certain
symmetries. We therefore need to use only three independent parameters to
parametrize the self-energy, 
\begin{eqnarray}
\Sigma &=& \sigma_3 \otimes \left( \begin{array}{ccc} \Sigma_1 & 0 & 0 \\ 0 &
    \Sigma_1 & 0 \\ 0 & 0 & \Sigma_2 \end{array}   \right) 
+ i\sigma_2 \otimes \left( \begin{array}{ccc} 0 & \Sigma_3 & 0 \\ -\Sigma_3 &
    0 & 0 \\ 0 & 0 & 0 \end{array}   \right) ,
\nonumber
\\
\label{eq:ansatz}
\end{eqnarray}
with $\sigma_2$ and $\sigma_3$ denoting the second and the third Pauli matrices in 
the Nambu space. Using this matrix, it is possible to solve selfconsistency relations
(\ref{eq:selfcons_equations}) numerically, and we can also calculate the
energy, 
\begin{equation}
E (g,\Delta, \rho_3 ) = K_{12}+K_3 + U N D_T \;, \label{eq:energy}
\end{equation}
where $N$ is the number of lattice sites and we explicitely indicated all implicit variables, with 
respect to which the energy must be optimized. The kinetic energy of the first two colors is
\begin{equation}
K_{12} = 2 \sum_\k \epsilon_\k  \frac{[1- \Sigma_1/(1+g) - f_\k
  \Sigma_3/(1+g)]^2}{
(1-\Sigma_1)(1+f_\k^2)},
\end{equation}
where $f_\k = \frac{\theta_\k - \Sigma_3}{1-\Sigma_1}$ is a renormalized
occupation number, with 
$\theta_\k = \frac{1}{\Delta} \left( \epsilon_\k - \mu_{12} + \sqrt{(\epsilon_\k - \mu_{12})^2 + \Delta^2} \right) $. The kinetic energy of the third color can be expressed as 
\begin{equation}
K_3 = \frac{[1- \Sigma_2/(1+g)]^2}{1-\Sigma_2} K_0 \;,
\end{equation}
where $K_0 = \sum_{\epsilon_\k < \mu_3} \epsilon_\k$
is the noninteracting kinetic energy of the atoms in channel 3.
Finally, the full double occupancy is given by
\begin{equation}
D_T = \frac{ 2 \Sigma_1 + \Sigma_2  - \sqrt{ (g^2 - 1) (\Sigma_1^2 +
    \Sigma_3^2) \Sigma_2}}{1 - g^2 - \sqrt{ (g^2 - 1) (\Sigma_1^2 +
    \Sigma_3^2) \Sigma_2} } . \label{eq:df}
\end{equation}

Notice that the above momentum sums as well as the summations included in the 
self-consistency equation contain terms that depend on the momentum only
through the single particle energy $\epsilon_\k$. It is therefore 
possible  to turn the momentum sums into energy integrals and
convert all these expressions into relatively simple self-consistent 
integral equations, involving only the density of states $D(\epsilon)$, 
which becomes the Gaussian on an infinite-dimensional cubic lattice, 
\begin{equation}
D(\epsilon) = \frac{1}{\sqrt{\pi t^\ast}} e^{-\epsilon^2/(t^\ast)^2}. 
\end{equation}
For high dimensional hypercubic lattices, the qualitative features of the densities of states are the same as in infinite dimensions. In this regard three dimension is probably close enough to infinite dimensions, and our results should hold away from half filling. One- and two-dimensional systems should be investigated by different methods since there spatial fluctuations and van Hove singularities will play a much more important role. Nevertheless, even in one dimension, recent Bethe ansatz\cite{trion-1D-BA} and density matrix renormalization group\cite{trion-1D-DMRG} calculations seem to support the results presented here and in Ref.~\onlinecite{su3-results-prl}.

\subsection{Numerical Results}

Our numerical procedure is as follows. First, we fix the 
filling factor $\rho$ and the interaction strength $U$. For a given set of variational 
parameters $g$, $\Delta$, and $\rho_3$, we use the ansatz 
(\ref{eq:ansatz}) and solve the selfconsistency relations
[Eq.~(\ref{eq:selfcons_equations})] by iteration.\cite{true_numerical} Once $\Sigma$ at hand, we can
compute the expectation value of the energy $E(g,\Delta,\rho_3)$ using Eqs.~(\ref{eq:energy})-(\ref{eq:df}). This
function can then be minimized numerically to find the optimum values for $g$, $\Delta$, and $\rho_3$.

\begin{figure}[tp]
\centering
\includegraphics[width=6cm,clip,angle=270]{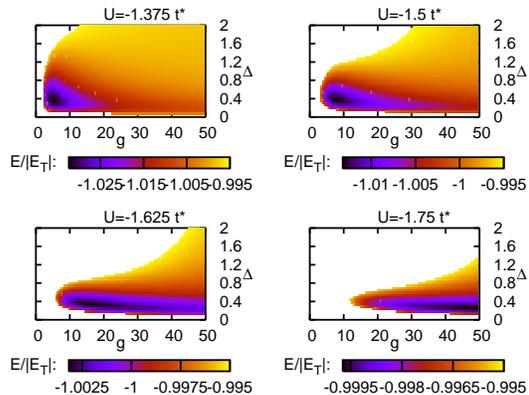}

\caption{\label{fig:landscape} (Color online) Variational energy $E(g,\Delta)$ compared to the energy of the trionic state, $E_T = 3 U N \rho$. Only regions of $E/|E_T| < -0.995$ are shown for different values of the interaction strength and $\rho_1=\rho_2=\rho_3=1/3$. }
\end{figure}

\begin{figure}[bp]
\centering
\includegraphics[width=7.5cm,clip]{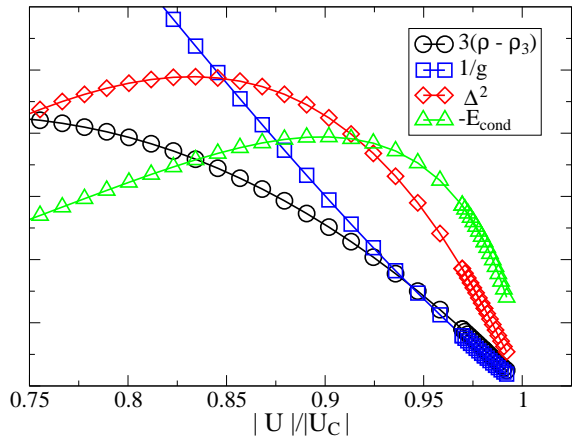}
\caption{\label{fig:secondorder} (Color online) Different ground state properties of the SU(3) attractive Hubbard model for $\rho=1/3$. For $|U|>|U_C| \approx 1.774 t^\ast$, the Gutzwiller parameter $g$ diverges, while the superfluid order parameter $\Delta$ and the condensation energy $E_{\rm cond}$ vanish. Energy is measured in units of $t^\ast$.}

\end{figure}

Typical results for a given density $\rho_1=\rho_2=\rho_3= 1/3$ are shown 
in Fig.~\ref{fig:landscape},  where the energy landscape $E(g,\Delta)$ is shown as a function of $g$ and $\Delta$ for various values of the 
coupling $U$. For small values of $|U|$, the overall energy minimum
is located at some finite value of the correlation parameter 
$g>1$ and gap $\Delta$. This energy minimum is shifted to larger and larger
values of $g$ as we increase $|U|$. The optimum value of 
 $\Delta$ initially increases with $|U|$, but at larger values of 
$|U|$, it starts to decrease again, the energy minimum 
is getting shallower and shallower and, eventually, it shifts to $g=\infty$ and $\Delta=0$ as 
$U$ approaches a critical value $U_C$. This behavior of the optimum values of 
$g$ and $\Delta$ is similar to the one shown in Fig.~\ref{fig:secondorder}.  Within numerical
accuracy, the gap vanishes continuously at $U=U_C$, which
is characteristic of a {\em second order} phase transition, i.e., a quantum
critical point. The correlation parameter $g$ precisely diverges at the same
value of $U$, implying that for larger values of $|U|$ the optimum ground state 
is a purely trionic state with no superconducting order.
As we discuss below, this scenario carries over for all
filling factors $0.0333<\rho<0.4833$ that we investigated, 
where  there is always an interaction strength $U_C(\rho)$ at which the Gutzwiller 
parameter $g$ diverges.

\begin{figure}[tp]
\centering
\includegraphics[width=7.5cm,clip]{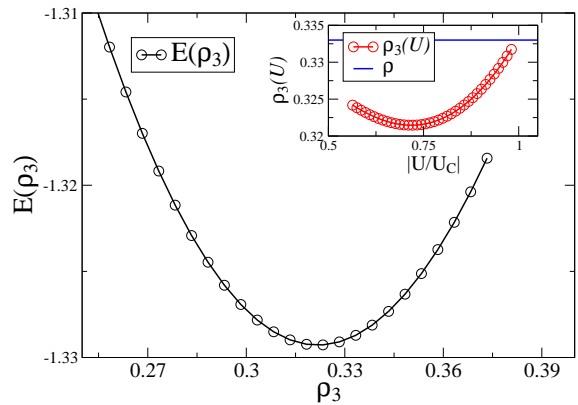}

\caption{\label{fig:n_3_dependence} (Color online) One can gain energy by transferring particles with color 3 to the superfluid channel: the energy minimum occurs for $\rho_3 < \rho$. The inset shows the optimal $\rho_3(U)$ for $\rho=1$. Energy is measured in units of $t^\ast$. }
\end{figure}

In the previous calculation, we fixed all three densities to be the same. 
We have done this in the spirit that in an optical lattice,
the total number of particles is fixed for each color.
However, one can decrease the energy of the system further 
in the superconducting phase by letting $\rho_3$ vary and only fixing the total
filling fraction $\rho$. As shown in Fig.~\ref{fig:n_3_dependence},
the optimal value of the average occupation of the third color is slightly less than the filling: $\rho_3 < \rho$. However, as shown in the inset
of Fig.~\ref{fig:n_3_dependence}, $\rho_3$ approaches 
 $\rho$ as $U$ approaches $U_C$. This is easy to understand: The reason of the 
formation of an imbalance is that one can gain condensation energy by 
transferring particles from color 3 to colors 1 and 2. However, the driving force
to create this imbalance is the superconducing condensation energy. 
This energy does not coincide with $\Delta$, but is rather defined as
the difference between the the energy of the state with $\Delta \ne 0$ and  the  energy of the Gutzwiller-correlated 
state with $\Delta= 0$, $E_G^{FS}$ 
\begin{equation}
E_{\rm cond} = E - E_G^{FS}.
\end{equation}
This energy is also shown in Fig.~\ref{fig:secondorder} and  it 
vanishes at the critical point $U=U_C$ as well. Therefore, 
the induced charge imbalance must also disappear there, in agreement with our numerical findings. 


In a system with SU(3) symmetry, this finding implies that, although the 
total number of particles is fixed for a given color, one can decrease the 
energy of the system by segregation,\cite{su3-results-prl,segregation_Cherng} i.e., by forming superconducting domains with unequal numbers of particles and orienting the order parameters to point in 
them in different directions, as we already sketched in Fig.~\ref{fig:sketch}. 
Of course, this is only possible in a large enough system, where the domain wall
energy is compensated by the overall condensation energy gain. 
Interestingly, the tendency to generate $\rho_3\ne \rho$ can also be viewed as the appearance of a secondary ferromagnetic order parameter, 
\begin{equation}
m^a\equiv \langle \hat c^\dagger \lambda^a c\rangle\ne 0\;.
\end{equation}
Therefore we conclude that this superconducting state is also a ferromagnet in
the SU(3) language.

\subsection{Breaking the SU(3) symmetry}

So far, we assumed that the scattering lengths are the same in all three
scattering channels. In most systems, however, the three 
scattering lengths are not equal and can vary with an external magnetic
field.  As we proposed in Ref.~\onlinecite{su3-results-prl}, a possible
candidate  for realizing the trionic state is $^6$Li. In this system, the magnetic field dependence of the 
three scattering lengths has been experimentally determined. 
\cite{feshbach-Li} According to the experimental results of Ref.~\onlinecite{feshbach-Li}, the interaction strengths can be approximated in the magnetic field region of 60 -- 120 mT as
\begin{equation}
U_{\beta\gamma} \approx U_0^{\beta\gamma}  \left[1 +\frac{\Delta^{\beta\gamma}}{B - B_0^{\beta\gamma}}\right]\left[1 +\alpha^{\beta\gamma} (B - B_0^{\beta\gamma})\right] \;, \label{eq:ualphabeta}
\end{equation}
where the parameters $B_0$, $\Delta$, and $\alpha$ in this equation can be
taken from Ref.~\onlinecite{feshbach-Li}, and $U_0^{12} \equiv U_0$, $U_0^{13} = 1.23\; U_0$, and $U_0^{23} = 1.06\; U_0$. The interaction amplitude $U_0=U_0^{12}$ can
be changed by tuning the potential depth.
For $B\gg B_r \equiv B_0^{12} = 83.41 {\rm mT} $, 
all three interactions become approximately 
the same, and are negative, $U_{\alpha\beta} <0$. On the other hand, if we 
increase the interaction strength by approaching the 
Feshbach resonance at $B_r$, then the interaction becomes rather anisotropic in color space. 

Experimentally, there are thus several ways to drive the system close to
the phase transition regime, $t^\ast/|U_0| \sim 1$. The first possibility is to change the intensity of the laser beams, and thereby tune mostly the
hopping amplitude $t^\ast$. In this case, one can apply a large magnetic field in which the interaction is almost SU(3) symmetrical, although it is somewhat 
stronger in channel 12 than in the others. The other possibility is to tune the ratio $t^*/|U_0| \sim 1$ by changing the magnetic field and approaching a Feshbach resonance. In this second case, the interactions can strongly break SU(3) symmetry.

In both cases, the anisotropic interaction has an important effect: it locks the
superconducting order parameter so that only $\Delta_{12}\ne 0$. The color superfluid phase thus becomes a more or less standard $U(1)$ superfluid with an additional decoupled Fermi liquid.  However, 
as we show, the trionic phase transition can survive in this anisotropic limit. 

\begin{figure}[tp]
\centering
\includegraphics[width=7.5cm,clip]{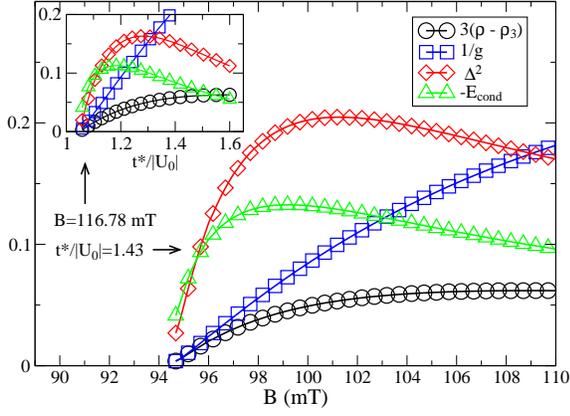}
\caption{ \label{fig:changeU0}  (Color online) Ground state properties of the Hamiltonian
(\ref{eq:H}) with the interaction strengths (\ref{eq:ualphabeta}) for $t^\ast/|U_0| = 1.43$ as a function of $B$ for $\rho=1/3$ filling, $U_0=U_0^{12}$. This describes a cold atomic system where three hyperfine states of $^6$Li are trapped and the magnetic field is tuned. The inset shows the same quantities with the magnetic field fixed at $B = 116.78 {\rm mT}$ as a function of $t^\ast/|U_0|$. Energy is measured in units of $t^\ast$.} 
\end{figure}

Let us now investigate the superfluid - trion transition, assuming that $ B > B_r$. In this region, all interactions are attractive, $U_{\alpha\beta} < 0$. Perturbation theory tells us that
the weak-coupling ground state is the color superfluid state
\cite{su3-smallU} formed in the channel 12, and in the strong coupling limit the ground state is a trionic state. Thus, we can use the same type of ansatz as in the SU(3)
case. Since the structure of the effective theory is uniquely determined by
the Gutzwiller ansatz state, the effective action, the Ward identities and
the self-consistency relations for the local effective action remain the same as before. The only difference is in the expression of the variational energy that we need to minimize,
\begin{eqnarray}
E(g,\Delta,\rho_3) = K + \sum_i [ U_{12} \langle \hat n_{i1} \hat
n_{i2}\rangle_G + U_{13} \langle \hat n_{i1} \hat n_{i3}\rangle_G \nonumber \\+ U_{23}
\langle \hat n_{i2} \hat n_{i3}\rangle_G   ] \;. 
\nonumber 
\end{eqnarray}
The previously used numerical procedure can be modified for this case, 
and we find that the quantum phase transition persists even with the
anisotropy in the interaction strengths. 

The anisotropy in $U_{\alpha\beta}$ has an important secondary effect: it induces somewhat different chemical potentials in channels 1 and 2, i.e., an SU(3) ``Zeeman field''. Deep in  the superfluid phase, this chemical potential difference is not sufficient to create a charge imbalance, $\rho_1 \ne\rho_2$, since breaking Cooper pairs requires a finite energy.\cite{Sarma} 
In principle, close to $U_C$, $\Delta_{12}\to0$, and therefore, the chemical potential difference could create a first order transition to the trionic state.
This would invalidate our restriction $\rho_1 = \rho_2$ in our calculations. However, at the critical point trionic correlations also diverge, $g\to \infty$, and therefore the effects of this Zeeman field are suppressed. In fact, within the Gutzwiller approach, the phase transition seems to be continuous even for anisotropic interactions (see Fig.~\ref{fig:changeU0}). In finite dimensions, on the other hand, we cannot exclude a first order transition.

In Fig. \ref{fig:changeU0},
we show different physical quantities computed for the ground state as a function of 
$ B$ at $\rho=1/3$ and $t^*/|U_0| = 1.43$.
The observed phase transition occurs at $B \approx 94.5 {\rm mT}$, and all parameters behave similarly to as in the SU(3) symmetrical case. 


\section{Analogy with quantum chromodynamics}
\label{sec:qcd}

The phase transition we found  is essentially the analogue of the color superconductor - baryon phase transition,\cite{qcd_ref} which is believed to occur in QCD. To make this  analogy with QCD clearer, let us 
rewrite  the partition function of the SU(3)-symmetrical Hamiltonian in a path integral form, 
\begin{eqnarray}
Z &=& \int {\cal D} \bar c \; {\cal D} c \; e^{- \int d\tau \;{\cal S}[ \bar
  c_{i\alpha}(\tau), c_{i\alpha}(\tau) ]}, \nonumber \\
{\cal S}&=&  \sum_{ij \alpha} \bar c_{i\alpha}(\tau){\cal G}^{-1}_{ij,\alpha}(\tau)
c_{j\alpha}(\tau) \nonumber \\ 
&&+ \frac U 2  \sum_i \sum_{\alpha  \beta} n_{i\alpha}(\tau) 
n_{i\beta}(\tau), 
\end{eqnarray}
where ${\cal G}^{-1}_{ij,\alpha} = \delta_{ij} (-\partial_\tau - \mu_\alpha) - t_{ij}$ 
denotes the inverse imaginary time Green's function for the non-interacting
atoms and $t_{ij}$ the hopping between sites $i$ and $j$. 

We can now decouple the interaction term by using the relation 
\begin{equation}
\sum_{\alpha \beta} n_{i\alpha} n_{i\beta} = - \frac{3}{8}  \sum_a \left( \bar
  c_{i} \lambda^a c_i 
\right)^2 ,
\end{equation} 
directly following from the SU(3) identity, 
\begin{equation}
\sum_a \lambda^a_{\alpha \beta} \lambda^a_{\gamma \kappa} = 2 \delta_{\alpha \kappa} \delta_{\gamma \beta} - \frac{2}{3} \delta_{\alpha \beta} \delta_{\gamma \kappa}\;,
\end{equation}
and then performing a Hubbard--Stratonovich transformation: 
\begin{eqnarray}
&{\rm exp} & \left(-U/2 \sum_{\alpha \beta} n_{i\alpha} n_{i\beta} \right) \nonumber \\
&{\rm exp} & \left(- \frac{3}{16} \vert U \vert \sum_a \left(  \bar c_{i} \lambda^a c_{i} \right)^2  \right)= \nonumber \\
& = C & \int \left(\prod_a dA^a_i\right)  \; {\rm exp}\left[ -  \sum_a (A_i^a)^2 - i \;\gamma
\sum_a A^a_i\; (\bar c_{i} \lambda^a c_{i})  \right], \nonumber
\end{eqnarray}
at every lattice and imaginary time point. Here we have suppressed the imaginary
time and color indices of the Grassmann variables $\bar c_{i}(\tau)$ 
and the  ``gluon field'', $A_i^a(\tau)$, and we introduced the coupling 
 $\gamma = \sqrt{3|U|}/2$ between the gluons and the fermions.

Using this transformation we can thus express the partition function as
\begin{eqnarray}
Z &=&\int {\cal D} A \;{\cal D} \bar c \; {\cal D} c \; {\rm exp}\left[- \int d\tau \;{\cal S}_{gl}[ \bar
  c_{i\alpha}(\tau), c_{i\alpha}(\tau), A^a_i (\tau)]\right] ,
\nonumber
\\
{\cal S}_{gl} &=& \sum_{ij\alpha} \bar c_{i}\; {\cal G}^{-1}_{ij}
c_{j} +
\sum_{i,a}(A_i^a)^2  + i\;\gamma\; \sum_{i,a} A^a_i  \;(\bar c_{i}\lambda^a c_{i}) .
\nonumber
\end{eqnarray}
In this language, the attractive interaction  is mediated by real bosonic fields, 
``gluons'', which are ultimately responsible for the formation of the trionic
(``baryonic'') phase.  Note, however, that in our case, gluons do not have 
a dynamics on their own, rather they are just massive Hubbard-Stratonovich fields. 
As a result, the mediated interaction is short-ranged, and there is 
no confinement.

In fact, our phase diagram  parallels the famous QCD phase diagram. 
In QCD, however, one changes the ratio of kinetic energy vs interaction energies 
through changing the chemical potential, while in our case, the hopping parameter
(i.e., the mass of the Fermions) is changed. Therefore, large chemical
potentials in the QCD correspond in our case to large values of 
$t^\ast$ (i.e., small $|U|$), while small chemical potentials 
in QCD correspond to  small $t^\ast$  (large $|U|$). Another important 
difference is the order of the phase transition: In our case, the two phases are
separated by a quantum critical point, while in QCD, the transition is of first
order, due to the long-ranged interactions.

Moreover, the superconducting phase we have is somewhat different from the one 
occurring in QCD, since cold atoms in optical lattices have only a color
quantum  number, while quarks in QCD have additional flavor degrees
of freedom. The color superfluid emerging
in our case has a nontrivial SU(3) color content and is
analogous to the color superconducting phase in two-flavor
QCD where only two flavors of light quarks are considered
.\cite{qcd_ref_2} In the alternative theoretical scenario of three-flavor
QCD, the superconducting state is expected to be color-flavor
locked.\cite{qcd_ref_3}

\section{Conclusions}
\label{sec:conclusions}

In this paper, we investigated a system of fermionic cold atoms with three internal
quantum states and attractive interactions among them.  We have shown  by
heuristic arguments as well as by a detailed variational calculation 
in the limit of $d \to \infty$-dimensions that this system shows a quantum phase transition 
from a correlated superconducting state to a trionic state, where 
fermions form three-body bound states and the superconducting order
disappears. 

We also showed that in the case of SU(3) symmetry, ferromagnetic ordering appears as a secondary order parameter. In a closed system of atoms, where the number of atoms is conserved for every color, this implies segregation, i.e., the
formation of domains where the color of the atoms forming the superconducting state changes from domain to domain and the density of the atoms involved in the superconducting order is slightly higher than that of the unpaired ones.

We have also demonstrated that the phase transition is robust
against breaking the SU(3) symmetry of the interaction. This symmetry breaking 
influences the structure of the superconducting state in that it pins the
order parameter to the most strongly interacting channels, but does not
influence the existence of the phase transition itself.

We have not studied, however, the case where an unequal number of atoms is
loaded for the three hyperfine states. This situation has been experimentally studied for a two-component $^6$Li system.\cite{two-comp-Li} There, a segregated superconducting state has been observed in the center of the trap,
where the two densities become equal in order to form Cooper pairs, 
while at the edge of the trap, spin densities are different. 
A similar scenario is expected for a slight imbalance of the three densities
in our case. It is, however, also possible that for SU(3) symmetrical 
interactions, three different superconducting domains of unequal
size can form for large enough traps. Formation of a Fulde--Ferrel--Larkin--Ovchinnikov state cannot 
be excluded either, although this state has not been experimentally observed .\cite{two-comp-Li}

Here, we should also remark that we solved the Gutzwiller problem only in $d=\infty$ dimensions, where it predicts a phase transition. Although the phase transition survives in finite dimensions too, there we do not know if the variational Gutzwiller wave function displays a phase transition or not. As long as $g$ remains finite, one has a step in the momentum distribution at the Fermi energy and the system can gain energy by forming a superconductor. Therefore, to have a phase transition, the parameter $g$ must diverge at a finite critical value of $U$. 
This certainly happens within the local approximation (Gutzwiller approximation) in finite dimensions too, and one can argue that this should also happen if one could evaluate the finite-dimensional Gutzwiller expectation values exactly. Unfortunately, restricted Monte Carlo calculations would be needed to give a definite answer to this question.

Finally, let us briefly discuss how the transition from a superfluid to a trionic state could be experimentally detected. One way to observe the condensate is by detecting vortices. In a perfectly SU(3)-symmetrical
system there are no vortices, because any vortex can be twisted away due to the large internal symmetry of the order parameter.\cite{su3-smallU} However, if one
breaks the SU(3) symmetry down to U(1) then the superconducting phase will
contain usual vortices that can be relatively easily observed optically.\cite{vortex-optical-detection} 
In case of $^6$Li, e.g., this can be achieved by approaching the 
Feshbach resonance at  $B_r=83.41$ mT from the high-field side, 
and thereby having a condensate in channel 12. Then our prediction is 
that the superfluid state and thus the vortices disappear for large enough 
$|U_{\alpha\beta}|/t^\ast$. This ratio can be tuned by either changing the 
laser beam intensities and thereby the amplitude of the optical lattice 
or by changing the external magnetic field. A further consequence of 
having $U_{13}\ne U_{23}$ is that the interaction induces a chemical potential 
difference for channels 1 and 2. Although we do not see a sign of it within
the infinite-dimensional Gutzwiller approach, this may well render the phase
transition first order in finite dimensions.

Pinning the order parameter to a single channel has a further advantage. 
Since one knows that the order parameter is simply in channel $\Delta_{12}$, one can 
use the method of Ref.~\onlinecite{projective-measurement} to measure $\Delta_{12}$ by sweeping
the magnetic field through the Feshbach resonance at $B_r=83.41 {\rm mT}$
and detecting the momentum distribution of the molecules thus 
formed. This method would enable one to detect 
$|\Delta|$  as a function of $|U_{\alpha\beta}|/t^\ast$ and verify the
existence of a critical value where it goes to zero. 

Stability of the three-color $^6$Li system may also be an important
issue. High magnetic fields will probably considerably stabilize the 
three-color condensate, but using mixtures of different fermionic atoms 
with all scattering lengths being negative may also be a possibility. 
In such a composite system, other interesting phenomena could also 
take place due to the difference in atomic masses. 

\section*{Acknowledgments}

We would like to thank C. Honerkamp for fruitful discussions and exchange of ideas. We also thank I. Bloch, E. Demler, D. Rischke, P. Zoller, and M. Zwierlein for useful discussions and comments. This research has  been supported by Hungarian OTKA under Grants  No. NF061726, No. T046303, No. T049571, and No. NI70594, and by the German Science Foundation (DFG) Collaborative Research Center (SFB TRR 49). 
G. Z. would like to thank the CAS, Oslo, where part of this work has been completed.

\appendix

\section{Ward identities}
\label{app:identities}

Here, we connect certain expectation values with the self energy $S$ defined by
Eq.~(\ref{eq:def_selfenergy}) to express the kinetic energy in
terms of the self-energy itself. First, we shift the integration variables of
$\Gamma$ in Eq.~(\ref{eq:def_gamma}) following 
Eq.~(\ref{eq:def_trf}). This transformation automatically transforms the
conjugate fields by 
\begin{equation} 
\bar \Psi_i \to \bar \Phi_i = \bar \Psi_i + \lambda \sum_p \bar I_p
  \tau_3 D^0_{pi}.
\end{equation}
Using these, the linear and quadratic terms become
\begin{eqnarray}
\frac{1}{2} \sum_{ij} \bar \Psi_i D^{0 -1}_{ij} \Psi_j + \sum_i \bar I_i
\tau_3 \Psi_i  \nonumber 
\\
\phantom{n}= \frac{1}{2} \sum_{ij} \bar \Phi_i D^{0 -1}_{ij}  
\Phi_j + (1-\lambda) \sum_i \bar I_i \tau_3 \Phi_i 
\nonumber 
\\
\phantom{n}+ \frac{\lambda^2 - 2 \lambda}{2} \sum_{ij} \bar I_i \tau_3 D^0_{ij} \tau_3
I_j \;. 
\end{eqnarray}
Transforming the interaction term is more complicated. First, we transform the on-site density,
\begin{eqnarray}
\bar \Psi_i \tau_3 \Psi_i &=& \bar \Phi_i \tau_3 \Phi_i - 2\lambda \bar \Phi_i
\tau_3 \sum_m D^0_{im} \tau_3 I_m \nonumber  \\
&+& \lambda^2 \sum_{nm} \bar I_n \tau_3 D^0_{ni} \tau_3 D^0_{im} \tau_3 I_m \;, \label{eq:transformaction1}
\end{eqnarray}
where we used that $\bar \Phi_i \tau_3 D^0_{ij} \tau_3 I_j = \bar I_j \tau_3 D^0_{ji} \tau_3 \Phi_i$.
Here, we note that all terms behave like scalars with respect to both the
Grassmann algebra and the color matrices; and they commute with each
other. When performing the functional derivation, we will have to expand the
exponential in terms of $I$ to the order ${\cal O} (I^2)$. Thus, when
calculating the third power of Eq.~(\ref{eq:transformaction1}), we can drop
certain higher order terms. The interaction term  can then be expanded as
follows:
\begin{eqnarray}
\sum_r && (\bar \Psi_r \tau_3 \Psi_r)^3 \nonumber \\ 
&& \approx \sum_r (\bar \Phi_r \tau_3
\Phi_r)^3  - 6 \lambda \sum_{r,m} (\bar \Phi_r \tau_3 \Phi_r)^2 (\bar \Phi_r \tau_3 
M^{-1}_{rm} \tau_3 I_m) 
\nonumber 
\\ 
&&\phantom{n} 
+ 12 \lambda^2 \sum_{r,n,m} 
(\bar I_n \tau_3
M^{-1}_{nr} \tau_3 \Phi_r) (\bar \Phi_r \tau_3 \Phi_r) 
(\bar \Phi_r \tau_3 M^{-1}_{rm} \tau_3 I_m) \nonumber 
\\ 
&&\phantom{n} + 3 \lambda^2 \sum_{r,n,m} 
(\bar \Phi_r \tau_3 \Phi_r)^2 (\bar I_n \tau_3 M^{-1}_{nr} \tau_3 M^{-1}_{rm}
\tau_3 I_m) + \cdots\;.
\nonumber
\end{eqnarray}
Now, we can expand the exponential in the definition of the generating
functional.
Introducing the coupling $u=g^2-1$ of the effective theory, we obtain
the following long expression
\begin{widetext}
\begin{eqnarray}
\Gamma &=& \ln \int {\cal D} \bar \eta {\cal D} \eta \; e^{- {\cal S} + \sum_i \bar I_i \tau_3 \Psi_i } \approx \nonumber \\
&\approx& \ln \int {\cal D} \bar a {\cal D} a \;  e^{- {\cal S}} \left(1 +  \frac{\lambda^2 -2\lambda}{2} \sum_{rr'} \bar I_r \tau_3 M^{-1}_{rr'} \tau_3 I_{r'}  +\frac{1}{2} \lambda^2 u \sum_{rr'r"} (\bar I_{r'} \tau_3 M^{-1}_{r'r} \tau_3 n^F_r \Phi_r)(\bar \Phi_r \tau_3 M^{-1}_{rr"} \tau_3 I_{r"}) \right. \nonumber \\ 
&&+ \frac{1}{2} \lambda^2 u \sum_{rr'r"} (\bar I_{r'} \tau_3 M^{-1}_{r'r} d^F_r \tau_3 M^{-1}_{rr"} \tau_3 I_{r"})   + \frac{1}{2} (1-\lambda)^2 \sum_{rr'} (\bar I_r \tau_3 \Phi_r)( \bar \Phi_{r'} \tau_3 I_{r'}) \nonumber \\
&& + \frac{1}{2} \lambda^2 u^2 \sum_{rr'pp'} (\bar I_{r'} \tau_3 M^{-1}_{r'r} \tau_3 \Phi_r d^F_r)( d^F_p \bar \Phi_p \tau_3 M^{-1}_{pp'} \tau_3 I_{p'}) \nonumber \\
&&  \left. -\frac{1}{2}\lambda(1-\lambda) u \sum_{rr'r"} (\bar I_r \tau_3 \Phi_r)(d^F_{r'} \Phi_{r'} \tau_3 M^{-1}_{r'r"} \tau_3 I_{r"}) - \frac{1}{2}\lambda(1-\lambda) u \sum_{rr'r"} (\bar I_r \tau_3
  M^{-1}_{rr'} \tau_3 d^F_{r'} \Phi_{r'} )(\bar \Phi_r" \tau_3 I_r")  \right)
\;, \nonumber \\
&&
\end{eqnarray}
where $a,\bar a$ are the transformed Grassmann fields, which correspond to $\Phi$.
Now one can perform the functional derivation to get the dressed propagator,
\begin{eqnarray}
D_{ij} &=& \left( \begin{array}{cc} \frac{\delta^2 \Gamma}{ \delta \bar J_i \delta J_j } & -\frac{\delta^2 \Gamma}{ \delta \bar J_i \delta \bar J_j } \\ -\frac{\delta^2 \Gamma}{ \delta J_i \delta J_j } & \frac{\delta^2 \Gamma}{ \delta J_i \delta \bar J_j } \end{array}  \right) \nonumber \\ 
&=& \tau_3 \frac{\delta^2 \Gamma}{ \delta \bar I \delta I } \tau_3 = (2\lambda - \lambda^2) D^0_{ij} - (1-\lambda)^2 \langle \Phi_i \bar \Phi_j \rangle_{\cal S} -  \lambda^2 u \sum_r D^0_{ir} \tau_3 \langle n^F_r \Phi_r \bar\Phi_r \rangle_{\cal S} \tau_3 D^0_{rj} \nonumber \\
&& - \lambda^2 u \sum_r D^0_{ir} \tau_3 \langle d^F_r \rangle_{\cal S} \tau_3 D^0_{rj}  - \lambda^2 u^2 \sum_{rr'} D^0_{ir} \tau_3 \langle \Phi_r d^F_r d^F_{r'} \bar \Phi_{r'} \rangle_{\cal S} \tau_3 D^0_{r'j} \nonumber \\ 
&& + (1-\lambda)\lambda u \sum_r  \langle \Phi_i d^F_r \bar \Phi_r  \rangle_{\cal S} \tau_3 D^0_{rj} + (1-\lambda)\lambda u \sum_r D^0_{ir} \tau_3 \langle \Phi_r d^F_r \bar \Phi_j  \rangle_{\cal S} \;.
\end{eqnarray}
Using that $D = -\langle \Phi \bar \Phi \rangle_{\cal S}$, we can rewrite this
in the following form: 
\begin{eqnarray}
D_{ij} &=& D^0_{ij} - \frac{\lambda}{2-\lambda} u \sum_r D^0_{ir} \tau_3 \langle n^F_r \Phi_r \bar\Phi_r  \rangle_{\cal S} \tau_3 D^0_{rj} - \frac{\lambda}{2-\lambda} u \sum_r D^0_{ir} \tau_3 \langle d^F_r
\rangle_{\cal S} D^0_{rj} \nonumber \\
&& -\frac{\lambda}{2-\lambda} u^2
\sum_{rr'} D^0_{ir} \tau_3 \langle \Phi_r d^F_r d^F_{r'} \bar \Phi_{r'}
\rangle_{\cal S} \tau_3 D^0_{r'j} \nonumber \\ 
&& + \frac{1-\lambda}{2-\lambda} u \sum_r  \langle \Phi_i d^F_r \bar \Phi_r
\rangle_{\cal S} \tau_3 D^0_{rj} + \frac{1-\lambda}{2-\lambda} u \sum_r D^0_{ir} \tau_3 \langle \Phi_r d^F_r \bar
\Phi_j  \rangle_{\cal S}  .
\end{eqnarray}
\end{widetext}
Substituting $\lambda=1$ leads to
\begin{eqnarray}
D_{ij} &=& D^0_{ij}  - \nonumber \\
&& - u \sum_r D^0_{ir} \tau_3 \langle n^F_r \Phi_r \bar\Phi_r \rangle_{\cal S}
\tau_3 D^0_{rj} + \nonumber \\ && - u \sum_r D^0_{ir} \tau_3 \langle d^F_r
\rangle_{\cal S} D^0_{rj} + \nonumber \\ && - u^2 \sum_{rr'} D^0_{ir} \tau_3
\langle \Phi_r d^F_r d^F_{r'} \bar \Phi_{r'} \rangle_{\cal S} \tau_3 D^0_{r'j}, 
\end{eqnarray}
while setting  $\lambda=0$ we obtain  another interesting identity,
\begin{eqnarray}
D_{ij}&=& D^0_{ij} + \frac{u}{2} \sum_r  \langle \Phi_i d^F_r \bar \Phi_r   \rangle_{\cal S}
\tau_3 D^0_{rj}  \nonumber \\ 
&& +  \frac{u}{2} \sum_r D^0_{ir} \tau_3 \langle \Phi_r d^F_r \bar \Phi_j   \rangle_{\cal S} \;.
\end{eqnarray}
Compared these to the definition of the self-energy
Eq.~(\ref{eq:def_selfenergy}), one finds the identities
Eqs.~(\ref{eq:identities-1}) - (\ref{eq:identities-3}) and thus the expression of the kinetic energy in
terms of the self-energy. 

\section{Derivation of the functional form of the local action}
\label{app:localaction}

The local action is defined by
\begin{equation}
\frac{1}{Z_L} e^{-{\cal S}_L} = \frac{1}{Z} \int {\cal D}' \bar \eta {\cal D}'
\eta \; e^{-{\cal S}} \;,
\end{equation}
where the prime means integration over all Grassmann variables except for
those defined at the origin. We shall derive its functional form in $d=\infty$
dimensions. It is useful to collect the terms in the action into three groups, 
\begin{equation}
{\cal S} = {\cal S}_0 + {\cal S}' + {\cal S}^{(0)} ,  
\end{equation}
where
\begin{equation}
{\cal S}_0 = -\frac{1}{2} \bar \Psi_0 [D^{0 -1} ]_{00} \Psi_0 - (g^2-1) t_0
\end{equation}
contains only terms at the origin, 
\begin{equation} 
{\cal S}^{(0)} = -\frac{1}{2} \sum_{i,j \neq 0} \bar \Psi_i [D^{0 -1} ]_{ij} \Psi_j - (g^2-1) \sum_{i \neq 0}  t_i 
\end{equation}
is the effective action on a lattice without the origin, and
\begin{equation}
{\cal S}' = -\frac{1}{2} \sum_{i \neq 0} \left [ \bar \Psi_i [D^{0 -1} ]_{i0} \Psi_0 +  \bar \Psi_0 [D^{0 -1} ]_{0i} \Psi_i \right] 	
\end{equation}
are the terms which connect the lattice and the cavity. ${\cal S}'$ can be viewed as a generating functional 
term with the currents $\bar h_i \equiv \bar \Psi_0  [D^{0 -1} ]_{0i} $ and their adjungates.
Thus,
\begin{eqnarray}
&\ln& \int {\cal D}' \bar \eta {\cal D}' \eta \; e^{-{\cal S}^{(0)} - {\cal S}'} = \nonumber \\
&=& \sum_n \sum_{i_1 \dots i_n, j_1 \dots j_n \neq 0} \bar h_{i_1} \cdots \bar
h_{i_n} h_{j_1} \cdots h_{j_n} \nonumber \\ & \times& \langle \Psi_{i_1}
\cdots \Psi_{i_n} \bar \Psi_{j_1} \cdots \bar \Psi_{j_n} \rangle^{\rm conn}_{{\cal S}^{(0)} ,}
\end{eqnarray}
since only connected graphs are generated this way. 
Let us first restrict ourselves to sites $i_q$ and $j_q$ being only nearest neighbors, and
consider terms of  order $n$. If all site indices are different, then
summation gives a factor $d^{2n}$. Since each bare propagator $D^{0-1}$ is
proportional to $t \sim 1/\sqrt{d}$, they give a prefactor $\sim d^{-n}$. The
correlation function is connected, and thus the distance between external
sites is $|| i - j || \geq 2$, and the scaling of the hopping ensures
that the correlation function is 
\begin{equation}
 \langle \Psi_{i_1} \dots \Psi_{i_n} \bar \Psi_{j_1} \dots \bar \Psi_{j_n} \rangle^{\rm conn}_{{\cal S}^{(0)}} \leq {\rm const}/d^{2n-1} .
\end{equation}
Thus, these terms give a contribution of the order $1/d^{n-1}$. 

If two of the labels $\{i_q\}$ and $\{j_q\}$ are the
same, then the summation gives a factor $d^{2n-1}$, however, 
 connected diagrams scale at most $1/d^{2n-2}$, and the 
final result is again $\sim 1/d^{n-1}$. This argument can be generalized to 
any combination of identical labels. 
This means that only the quadratic terms $n=1$ survive the $d \to \infty$
limit,  which can be added to the quadratic term in ${\cal S}_0$, leading to
the bare propagator of  the local theory ${\cal D}_0$. The self-consistency
relations are required to determine its value.

This line of argumentation can also be extended to the case where $i_q$ and $j_q$ run over non-nearest neighbor sites. One only has to use the property that $(D^0)^{-1}_{ij}$ falls off faster than $1/d^{||i-j||/2}$, with $||i-j||$ the number of steps needed to reach the lattice site $j$ from site $i$.

\section{Diagonalization and symmetry properties of the local action}
\label{app:diagonal}

In this Appendix, we show that the  matrix $U$ that diagonalizes the 
local propagator ${\cal D}^{-1}_0$ has a special symplectic symmetry, 
which implies that one can evaluate the Green's functions by performing 
a variable transformation with this matrix.
 We first observe that the Nambu spinor $\Psi$ satisfies the relation
$\bar \Psi = \Psi^T\tau_1 $ where
$\tau_1 = \sigma_1 \otimes \delta_{\alpha \beta}$
and $\sigma_1$ is the
first Pauli matrix acting in the Nambu space. Therefore, 
we can  parametrize ${\cal D}^{-1}_0$ in the following way, 
\begin{equation}
	{\cal D}^{-1}_0 = \left( \begin{array}{cc} H  & A \\  
A^+ & - H^T \end{array}
\right) \;,
\end{equation}
where $H$ is Hermitian and $A$ is antisymmetric. Thus, the following identity
holds
\begin{equation}
	\tau_1 {\cal D}^{-1}_0 \tau_1 = -{\cal D}^{-1 *}_0 \;.
\end{equation}
This structure of ${\cal D}^{-1}_0$
enables us to say something about its
eigenvalues. Let us assume that $\underline{\mathbf v}_i$ is 
a right-hand side eigenvector of ${\cal D}^{-1}_0$ (a column vector), 
\begin{equation}
{\cal D}^{-1}_0 \underline{\mathbf v}_i = \epsilon_i \underline{\mathbf v}_i
\;.	
\end{equation}
Then, $\underline{\mathbf w}_i= \tau_1 \underline{\mathbf v}_i^*$ is also an
eigenvector, 
\begin{equation}
	{\cal D}^{-1}_0 \tau_1 \underline{\mathbf v}_i^* = \tau_1 (\tau_1 {\cal
  D}^{-1}_0 \tau_1 ) \underline{\mathbf v}_i^* = -\tau_1  {\cal D}^{-1 *}_0
\underline{\mathbf v}_i^* = - \epsilon_i \tau_1 \underline{\mathbf v}_i^* \;,
\end{equation}
since the eigenvalues are real. As a consequence, we can construct a unitary
matrix ${\cal U}^+$ as
\begin{equation}
	{\cal U}^+ = (\underline{\mathbf v}_1,\underline{\mathbf
  v}_2,\underline{\mathbf v}_3, \tau_1 \underline{\mathbf v}^*_1,\tau_1
\underline{\mathbf v}^*_2,\tau_1 \underline{\mathbf v}^*_3)
\end{equation} 
which transforms ${\cal D}^{-1}_0$ to a diagonal form, 
\begin{equation}
	{\cal U} {\cal D}^{-1}_0 {\cal U}^+ = \left( \begin{array}{cc} {\rm
        d}^{-1} & 0 \\ 0 & -{\rm d}^{-1}  \end{array} \right)  \;,   
\label{eq:diag}
\end{equation}
with ${\rm d}$ is a real diagonal matrix. 
 
Now, we can show or see that $\cal U$ is a well-defined transformation for
$\Psi$ in the sense that the identity 
$\bar \Psi = \Psi^T\tau_1 $ 
also holds for the transformed eigenvectors, $\Phi= {\cal U} 
\Psi,\bar \Phi = \bar \Psi {\cal U}^+$, and
\begin{equation}
	\bar \Phi = \Phi^T \tau_1 \;.
\label{eq:transposed}
\end{equation}
Note that this identity is necessary to carry out the path intergral in terms 
of the components of the  transformed field $\Phi$ as new independent variables. 
This equality is satisfied if 
\begin{equation}
	\bar \Psi {\cal U}^+ = \bar \Phi = \Phi^T \tau_1 = ({\cal U} \Psi)^T
\tau_1=\Psi^T {\cal U}^T \tau_1 , 
\end{equation}
i.e., if the matrix ${\cal U}$ satisfies the condition
\begin{equation}
	{\cal U}^+ = \tau_1 {\cal U}^T \tau_1 . \label{eq:well-definition}
\end{equation}
Equation (\ref{eq:well-definition}) can readily be verified by using the properties of the 
eigenvectors
\begin{eqnarray}
\tau_1 {\cal U}^T \tau_1 &=& \tau_1 (\underline{\mathbf v}^*_1,\underline{\mathbf v}^*_2,\underline{\mathbf v}^*_3, \tau_1 \underline{\mathbf v}_1,\tau_1 \underline{\mathbf v}_2,\tau_1 \underline{\mathbf v}_3) \tau_1 = \nonumber \\ &=&   \tau_1 (\tau_1 \underline{\mathbf v}_1,\tau_1 \underline{\mathbf v}_2,\tau_1 \underline{\mathbf v}_3, \underline{\mathbf v}^*_1,\underline{\mathbf v}^*_2,\underline{\mathbf v}^*_3) = \nonumber \\ &=& (\underline{\mathbf v}_1,\underline{\mathbf v}_2,\underline{\mathbf v}_3, \tau_1 \underline{\mathbf v}^*_1,\tau_1 \underline{\mathbf v}^*_2,\tau_1 \underline{\mathbf v}^*_3) = \nonumber \\ &=& {\cal U}^+ \;. 
\end{eqnarray}
 Relations [Eqs.~(\ref{eq:diag}) and (\ref{eq:transposed})] together with these 
imply that the Green's function of the local Green's
function can be computed by first transforming to the diagonal basis of 
${\cal D}^{-1}_0$ and then transforming back the Green's function to the 
original basis by using the transformation ${\cal U}$. 
Importantly, the interaction term of the local action
is invariant under the transformation $\Psi\to\Phi$. This follows simply from
the facts that the interaction can also be expressed as $\sim
\prod_{\mu=1}^6\Psi_\mu$ and that the determinant of $U$ is just 1.


\begin{thebibliography}{99}
\bibitem{Anderson} M. H. Anderson, J. R. Ensher, M. R. Matthews, C. E. Wieman,
  and E. A. Cornell, Science \textbf{269}, 198 (1995). 

\bibitem{Li-7} C. C. Bradley, C. A. Sackett, J. J. Tollett, and R. G. Hulet, Phys. Rev. Lett. \textbf{75}, 1687 (1995).

\bibitem{Na-23} K. B. Davis, M. -O. Mewes, M. R. Andrews, N. J. van Druten, D. S. Durfee, D. M. Kurn, and W. Ketterle, Phys. Rev. Lett. \textbf{75}, 3969 (1995).

\bibitem{Li-6} H. T. C. Stoof, M. Houbiers, C. A. Sackett, and R. G. Hulet, Phys. Rev. Lett. \textbf{76}, 10 (1996).

\bibitem{K-40} C. A. Regal, M. Greiner, and D. S. Jin, Phys. Rev. Lett. \textbf{92}, 040403 (2004).

\bibitem{Cs-133} D. Boiron, C. Triche, D. R. Meacher, P. Verkerk, and G. Grynberg, Phys. Rev. A \textbf{52}, R3425 (1995).

\bibitem{BEC_BCS} M. W. Zwierlein, C. A. Stan, C. H. Schunck, S. M. F. Raupach, A. J. Kerman, and W. Ketterle, Phys. Rev. Lett. \textbf{92}, 120403 (2004); M. Bartenstein, A. Altmeyer, S. Riedl, S. Jochim, C. Chin, J. Hecker Denschlag, and R. Grimm, \textit{ibid.} \textbf{92}, 120401 (2004).

\bibitem{bosonic_Mott} M. Greiner, O. Mandel, T. Esslinger, T. W. Hansch, and I. Bloch, Nature \textbf{415}, 39 (2002).

\bibitem{Hubbard_approx_Jaksch} D. Jaksch, C. Bruder, J. I. Cirac, C. W. Gardiner, and P. Zoller, Phys. Rev. Lett. \textbf{81}, 3108 (1998).

\bibitem{Hubbard_approx_Walter} W. Hofstetter, J. I. Cirac, P. Zoller, E. Demler, and M. D. Lukin, Phys. Rev. Lett. \textbf{89}, 220407 (2002).

\bibitem{Li7+Li6} A. G. Truscott, K. E. Strecker, W. I. McAlexander, G. B. Partridge, and R. G. Hulet, Science \textbf{291}, 2570 (2001).

\bibitem{evap_cooling} B. DeMarco, and D. S. Jin, Science \textbf{285}, 1703 (1999); Z. Hadzibabic, C. A. Stan, K. Dieckmann, S. Gupta, M. W. Zwierlein, A. G\"orlitz, and W. Ketterle, Phys. Rev. Lett. \textbf{88}, 160401 (2002).

\bibitem{Fermi_deg_opticallattice_Kohl} M. K\"ohl, H. Moritz, T. St\"oferle, K. G\"unther, and T. Esslinger, Phys. Rev. Lett. \textbf{94}, 080403 (2005).

\bibitem{Fermi_deg_opticallattice_Folling}  S. F\"olling, F. Gerbier, A. Widera, O. Mandel, T. Gericke, and I. Bloch, Nature \textbf{434}, 481 (2005).


\bibitem{su3-smallU} C. Honerkamp, and W. Hofstetter, Phys. Rev. Lett. \textbf{92}, 170403 (2004); C. Honerkamp and W. Hofstetter, Phys. Rev. B \textbf{70}, 094521 (2004).

\bibitem{su3-results-prl} A. Rapp, G. Zarand, C. Honerkamp, and W. Hofstetter, Phys. Rev. Lett. \textbf{98}, 160405 (2007 ).

\bibitem{threefermionproblem} T. Luu, and A. Schwenk, Phys. Rev. Lett. \textbf{98}, 103202 (2007).

\bibitem{threespeciessuperfluidity} H. Zhai, Phys. Rev. A \textbf{75}, 031603 (R) (2007).

\bibitem{Yin} J. Yin, Phys. Rep. \textbf{430}, 1 (2006).

\bibitem{feshbach-Li} M. Bartenstein, A. Altmeyer, S. Riedl, R. Geursen, S. Jochim, C. Chin, J. Hecker Denschlag, R. Grimm, A. Simoni, E. Tiesinga, C. J. Williams, and P. S. Julienne, Phys. Rev. Lett. \textbf{94}, 103201 (2005).


\bibitem{Subir} S. Powell, S. Sachdev, and H.P. B\"uchler, Phys. Rev. B {\bf 72}
  024534 (2005).

\bibitem{dmft-revmodphys} A. Georges, G. Kotliar, W. Krauth, and M. J. Rozenberg, Rev. Mod. Phys. \textbf{68}, 13 (1996).

\bibitem{multiband_gutzwiller} J. B\"unemann, F. Gebhard, and W. Weber, J. Phys.: Condens. Matter \textbf{9}, 7343 (1997); J. B\"unemann, W. Weber, and F. Gebhard, Phys. Rev. B \textbf{57}, 6896 (1998).

\bibitem{segregation_Cherng} R. W. Cherng, G. Refael, and E. Demler, Phys. Rev. Lett. \textbf{99}, 130406 (2007).


\bibitem{SU3-rotation}  This can be done without loss of generality, since in the
superfluid state one can always perform a suitable rotation
such that only the 12 component of the order parameter
$\Delta_{\alpha\beta} \sim \langle c_{i\alpha} c_{i\beta} \rangle$ is non-zero.


\bibitem{Negele} J. W. Negele and H. Orland, Quantum Many-particle Systems (Perseus Books, Reading, MA, 1998).

\bibitem{trion-1D-BA} X. W. Guan, M. T. Batchelor, C. Lee, H.-Q. Zhou, arXiv:0709.1763 (unpublished).

\bibitem{trion-1D-DMRG} S. Capponi, G. Roux, P. Lecheminant,
  P. Azaria, E. Boulat, and S. R. White, Phys. Rev. A \textbf{77}, 013624 (2008).


\bibitem{true_numerical} Using Eq.~(\ref{eq:dyson_local}), we can also express the self-consistency equations [Eq.~(\ref{eq:selfcons_equations})] in terms of the components of ${\cal D}_0$. For numerical stability reasons, we solved these equivalent equations rather than Eq.~(\ref{eq:selfcons_equations}) for $\Sigma$.Then, we determined $\Sigma$ from ${\cal D}_0$ using Eq.~(\ref{eq:dyson_local}).

\bibitem{Sarma} G. Sarma, J. Phys. Chem. Solids {\bf 24}, 1029 (1963).

\bibitem{qcd_ref} Z. Fodor and S. D. Katz, J. High Energy Phys. \textbf{03} 014 (2002).

\bibitem{qcd_ref_2} M. Alford, K. Rajagopal, and F. Wilczek, Phys. Lett. B \textbf{422}, 247 (1998).

\bibitem{qcd_ref_3} M. Alford, K. Rajagopal, and F. Wilczek, Nucl. Phys. B \textbf{537}, 443 (1999).

\bibitem{two-comp-Li} M. W. Zwierlein, A. Schirotzek, C. H. Schunck, and W. Ketterle, Science \textbf{311}, 492 (2006).

\bibitem{vortex-optical-detection} M. W. Zwierlein, J. R. Abo-Shaeer, 
A. Schirotzek, C. H. Schunck,  and W. Ketterle, Nature \textbf{435}, 1047 (2005).


\bibitem{projective-measurement} M. W. Zwierlein, C. H. Schunck,
  A. Schirotzek,  and W. Ketterle, Nature(London) \textbf{442}, 54 (2006). 





\end{thebibliography}
\end{document}